\documentclass[]{aa}
\usepackage{graphicx}
\usepackage{times}
\usepackage{natbib}
\usepackage{amssymb,amsmath}
\bibpunct{(}{)}{;}{a}{}{,} 
\graphicspath{{eps_dist/}}
\allowdisplaybreaks

\newcommand{\be}{ \begin {equation}}
\newcommand{\ee}{ \end {equation}}
\newcommand{\G}{ \mathcal{G}}
\newcommand{\degree}{\ensuremath{^\circ}}
\newcommand{\changed}[1]{#1}

\begin{document}

\titlerunning{Mean Motion Resonances and Stochastic Forcing}
\title{On the evolution of mean motion resonances
through stochastic forcing: 
Fast and slow 
libration modes and the
origin of HD128311}
\authorrunning{H. Rein \and J.C.B. Papaloizou}
\author{
	Hanno Rein
		\and
	John C. B. Papaloizou 
}
\institute{
	Department of Applied Mathematics and Theoretical Physics, University of Cambridge, Centre for Mathematical Sciences, Wilberforce Road, Cambridge CB3 0WA, United Kingdom \\ 
	\email{hr260@cam.ac.uk}
}

\date{Received November 11, 2008 - Recommended for publication in A\&A.} 

\abstract
{}
{We clarify 
the response of extrasolar planetary systems
in a 2:1 mean motion commensurability with masses ranging from  the super Jovian  range to the terrestrial range
to stochastic forcing that could result from protoplanetary disk turbulence. 
The behaviour of the different libration modes for a  wide range of system parameters
and  stochastic forcing magnitudes is investigated.
The growth of libration amplitudes  is parameterized as a function of the relevant \changed{physical} parameters.
The results are applied to provide an explanation of the configuration of the HD128311 system.
}
{We first develop an analytic model \changed{from first principles} without making the assumption that both eccentricities
are small. We also perform numerical N-body simulations with additional stochastic forcing terms to represent the effects of putative disk turbulence. }
{We isolate two distinct libration modes for the resonant angles. 
These react to stochastic forcing in a different way and become coupled
when the libration amplitudes are large.  Systems are quickly destabilized
by large  magnitudes of  stochastic forcing but some stability is imparted
should  systems undergo a net orbital migration. The slow mode, which mostly corresponds to
motion of the angle between the apsidal lines of the two planets, is converted to circulation
more readily than the fast mode which is associated with oscillations of the semi-major axes.
This mode is also vulnerable to the attainment of small eccentricities which causes
oscillations between periods of libration and circulation.   
 }
{ Stochastic forcing due  to disk turbulence may have played a role
in shaping the configurations of observed systems in mean motion resonance.
It naturally provides a mechanism  for accounting for the HD128311 system 
for which the fast mode librates and the slow mode is apparently 
near the borderline between libration and circulation. }


\maketitle

\section{Introduction}
Of the recently discovered \changed{335} extrasolar planets, at least 75 are in multiple planet systems \citep{exoplanet}. About 10\%  of these are in or very close to a resonant configuration where two planets show a mean motion commensurability, with four systems in or near a 2:1 resonance \citep{Udry07,Reipurth07}.

Resonant configurations can be established by dissipative forces acting on the planets which lead to convergent migration \citep[see for example][] {LeePeale01, SnellgrovePapaloizouNelson01}. An anomalous  effective kinematic viscosity~$\nu$ of order~$\sim 10^{-5},$ but with considerable uncertainty, has been  inferred from observations of accretion rates onto the central protostars. The magneto rotational instability (MRI) is thought to be responsible for this anomalous value of $\nu$ which is often characterized  using  the Shakura \& Sunyaev (1973) $\alpha =\nu/(c_s \Omega)$ parameter, with $c_s$  being the local sound speed and $2\pi/\Omega$ being
the orbital period. For the most likely situation of small or zero net magnetic flux,
recent  MHD simulations have indicated $\alpha\approx 10^{-2} - 10^{-3}$.  Even assuming
adequate ionization, this value is uncertain when realistic values of the actual transport coefficients are employed
because of numerical resolution issues \citep[see][]{FromangPapaloizou07, Fromangetal07}. 
Which parts of  protoplanetary disks are adequately ionized, or consitute a dead zone,
 is also an issue \citep[eg.][]{Gammie96}.
Because the level of MRI turbulence and associated density
fluctuations are very uncertain, we consider stochastic force amplitudes ranging over
several orders of magnitude.

The influence of stochastic forces resulting from  the gravitational
field produced by the density fluctuations associated with MRI turbulence
on migrating planets was explored first by \cite{NelsonPapaloizou04}.
They considered MRI simulations directly with the result that the simulation ran
only for a relatively small number of orbits.
To consider much longer evolution times, as done here, a simpler
parameterized model of the forcing is needed.

\changed{
Recent studies by   \citet{Adams08} and  \citet{LecoanetAdams2008}   have been made
in order to estimate  the  lifetimes of the  mean motion resonances in the GJ876~system
 when it is  perturbed by a sequence of  stochastic kicks. They found that resonances were disrupted
within  expected disk lifetimes for sufficiently large forcing.
They assume a fixed orbit for the outer planet and gave an interesting discussion 
of  the interactions of the two planets based on 
 a pendulum equation   with an additional ad hoc stochastic forcing term. 
The aim of the  work presented in this paper is to provide  a more complete study of the full 
set of celestial mechanics equations and to  describe the interplay between apsidal 
corotation and mean motion resonance  as well as to consider a wide range of planet masses and stochastic force amplitudes. 
In addition, we derive a  prescription for incorporating continuous stochastic 
forcing terms that allows for a general autocorrelation function
with associated correlation time.
 Their magnitudes are related to properties of the protoplanetary disk and  physical scaling laws are found.
 This is of particular importance as  the   factors that control 
 the strength of the stochastic forcing are not well constrained.
We also use our formalism to  estimate the stochastic diffusion rates of the orbital elements 
 from first principles.
}

\changed{ We further remark that
\cite{Adams08} gave a discussion of the diffusion of resonant angles that do not satisfy the d'Alambert condition \citep[see eg.][]{Hamilton94} which is equivalent to the requirement of rotational invariance.
Thus their appearrance  cannot be 
  straightforwardly connected to the basic equations or angles discussed in this paper unless
one assumes that the variation of the  longtitude of 
pericentre of one or both of the interacting planets can be neglected.
However, the system lifetimes derived from their numerical work is broadly consistent with ours
for the parameter regime they considered.  
}

\changed{
We find  that stochastic forcing readily produces systems in mean motion resonance with  broken apsidal corotation. 
An additional aim of this paper
is to use this feature  to construct  scenarios involving convergent
migration and stochastic forcing to account for the HD128311 system. 
}

The plan of the paper is as follows.
In section 2 we present the basic equations governing two interacting
planets subject to external stochastic forces.
We specialize to the case where the planets are either in or near a mean motion commensurability
and so retain only the two angle variables that 
vary on a time scale much longer than the  orbital one.
Considering the case where these angles undergo small amplitude librations,
we identify fast and slow libration modes, the former being associated with variations
in the semi-major axes and the latter with the angle between the apsidal lines of the two planets.
We go on to consider the effects of stochastic forcing arising from some external process
such as disk turbulence in  section 2.3 deriving the diffusion rates for the orbital
elements of a single planet and the growth rate of the libration amplitudes 
in the two planet case.

In section 3 we discuss the origin  and numerical implementation  of the stochastic forcing
and their operation in the single planet case. In section 4 we go on to consider the
stochastic forcing in the two planet case. We consider the conversion of the fast and slow
modes from libration to circulation for model systems of varying total
mass ratio and initial  eccentricities including GJ876. 
\changed{ The attainment of small eccentricities and coupling to the fast
mode results in conversion of the slow mode to circulation  before the fast mode.}
Investigations are carried out for stochastic diffusion rates, proportional to the mean square
stochastic force amplitude ranging over several orders of magnitude. The life
time of the resonant angles is found to be inversely proportional  to the diffusion rate
except in the case of systems with low total mass in the earth mass range.

In section 5 we exploit the tendency of the slow mode, related to the angle between
the apsidal lines, to  be driven to circulate while  the fast mode still librates
in stochastically forced systems, 
to make a model for the formation of the HD128311 system which may be in such a state and was not readily understood in terms of  convergent migration models for producing the commensurability. We combine the effects of such  migration and stochastic forcing, showing
that during the migration phase, while the librations tend to be stabilized, the slow mode
is readily converted to circulation while the fast mode continues to librate.
Good agreement is obtained with the somewhat uncertain observed orbital configuration.
Finally in section 6 we summarize and discuss our results.

\section{Basic Equations}
We begin by writing down the equations of motion for a single planet
moving in a fixed plane under a general Hamiltonian $H$ in the form
\citep[see e.g.][]{SnellgrovePapaloizouNelson01, PapaloizouResonances}
\begin{eqnarray}
\dot E &=& -n\frac{\partial H}{\partial \lambda}\label{eqnmo1}\\
\dot G &=& -\left(\frac{\partial H}{\partial \lambda}+\frac{\partial H}{\partial \varpi}\right)\\
\dot \lambda &=& \frac{\partial H}{\partial L} + n \frac{\partial H}{\partial E}\\
\dot \varpi &=& \frac{\partial H}{\partial L}.\label{eqnmo4}
\end{eqnarray} 
\changed{Here the angular momentum of the planet is $G$ and the energy is $E.$
For orbital motion around a central point  mass $M$ we have  }
\begin{eqnarray}
     G &=&  m\sqrt{\G Ma(1-e^2)} \hspace{0.5cm}  {\rm and} \\
     E &=& -{{\G Mm}\over{2a}},
\end{eqnarray}
\changed{where $\G$ is the gravitational constant, $a$ the semi-major axis and $e$ the eccentricity.}
The  mean longtitude is $\lambda = n (t-t_0) +\varpi ,$ 
 where $n$ the mean motion, with
$t_0$ being the time of periastron passage and $\varpi$ being the longitude of periastron.

\subsection{Additional Forcing of a Single Planet}\label{AdF}
In order to study the phenomena such as stochastic forcing we need to consider
the effects of an additional external force per unit mass ${\bf F}$ which may not be  described using  a Hamiltonian formalism.
However, as may be seen by considering  general coordinate transformations starting from a Cartesian representation,
the equations of motion are linear in the components of  ${\bf F}$.
Because of this we may determine them by considering forces of the form ${\bf F} = (F_x, F_y)$
for which the Cartesian components are constant.  Having
done this we may  then suppose that these  vary
with coordinates and time in an arbitrary  manner.
Following this procedure we note that when ${\bf F},$ as in the above form is constant, we may derive the equations
of motion by replacing  the original Hamiltonian  with a new Hamiltonian defined through
\begin{eqnarray}
H &\rightarrow& H - m\left( F_x\;x+F_y\;y\right) = H  - m\left(\mathbf{r}\cdot\mathbf{F}\right).
\end{eqnarray} 
The additional terms proportional to the components of ${\bf F}$ correspond to the Gaussian
form of the equations of motion \citep{BrouwerClemence1961}.

The  various derivatives involving  $\mathbf r$ can be calculated by elementary means and expressed in terms of $E, G, \lambda$ and $\varpi.$ 
One thus finds  additional contributions to the equations of motion
 (\ref{eqnmo1})~-~(\ref{eqnmo4}), indicated with a subscript $F,$ in the form
\begin{eqnarray}
\dot G_F &=&m\left( \frac{\partial }{\partial \lambda } + \frac{\partial }{\partial\varpi }   \right)
 \;\left(\mathbf{ r \cdot F}\right) =
 m\;\mathbf{\left( r\times F \right) \cdot \hat e}_z \label{feqnmo1}\\
\dot E_F &=& m n\frac{\partial }{\partial \lambda} \left( \mathbf{ r \cdot F} \right) = m\;(\mathbf {v \cdot F} )\label{feqnmo2} \\
 \dot \varpi_F &=& - m\frac{\partial }{\partial L} \left( \mathbf{ r \cdot F} \right) \ \ \ {\rm or \ \ equivalently} \\
\dot \varpi_F  &=& \frac{ \sqrt{(1-e^2)} }{n a\,e}  
 \left[F_\theta \left( 1 + \frac{1}{{1-e^2}} \frac {r}{a} \right)  \sin f
 - F_r  \cos f \right]  \label{feqnmo3}\\
\dot \lambda_F &=& - m\left(\frac{\partial }{\partial L}  
 + n  \frac{\partial}{\partial E} \right) \left( \mathbf{ r \cdot F} \right) \\
 &=& 
  \left( 1- \sqrt{1-e^2}\right) \dot \varpi_F 
 +\frac{2an}{\G M} \;(\mathbf{r\cdot F}) , \label{feqnmo4}
\end{eqnarray}
\changed{where the true anomaly $f$ is defined as the difference between the true longitude and the longitude of periastron, $f = \theta - \varpi.$ Note that from (\ref{feqnmo2})
we obtain}
\begin{eqnarray}
  \dot a_F \ \  = \ \ -{2 a \dot n\over 3 n} 
 &=& \frac {2( {F_r} e \sin f + {F_{\theta}}(1+ e \cos f))}
	{n\sqrt{1-e^2}}.\label{feqnmo6}
\end{eqnarray}
and from  (\ref{feqnmo1}) together with  (\ref{feqnmo2})
we obtain
\begin{eqnarray}
	\dot e_F &=& \frac{G(2 E \dot G + G \dot E) }{\G^2m^3M^2 e}.
\end{eqnarray}
In the limit $e \ll 1$ this becomes (ignoring terms $O(e)$ and smaller)
\begin{eqnarray}
	\dot e_F &=&  {F_r} \frac 1 {an} \sin f + {F_{\theta}} \frac 1{an} 2 \cos f.\label{feqnmo5}
\end{eqnarray}
 Furthermore in this limit we may replace $f$ by $f= \lambda -\varpi = n(t-t_0).$

{\changed{ We remark that the above formalism results 
in  equation (\ref{feqnmo3}) which
gives an expression for $\dot \varpi_F$ that indicates that this quantity diverges
for small $e$ as ${1/ e}.$ 
We comment that, as is well known,  this aspect results from the choice 
of coordinates used and is not associated with any actual singularity
or instability in the system.
This is readily seen if one uses $h=e\sin\varpi,$ and $k=e\cos\varpi$ as
dynamical variables rather than $e$ and $\varpi.$ The former
set behave like Cartesian coordinates, while the latter  set are the 
corresponding cylindrical polar coordinates. When the former set  are 
used,  potentially divergent terms $\propto 1/e$ do not appear.
 This can be seen from (\ref{feqnmo3}) and  (\ref{feqnmo5}) which give  
in the small $e$ limit}} 
\begin{eqnarray}
 \dot h_F &=& -{F_r} \frac 1 {an} \sin \lambda + {F_{\theta}} \frac 1{an} 2 \cos \lambda
\label{feqnmo7} \\
  \dot k_F &=&  {F_r} \frac 1 {an} \sin \lambda + {F_{\theta}} \frac 1{an} 2 \cos \lambda
.\label{feqnmo8}
\end{eqnarray}
\changed{Abrupt changes to $\varpi$ may occur when $h$ and $k$ pass through the
origin in the $(h,k)$ plane. But this is clearly just due to a coordinate
singularity rather than a problem with the physical system
which changes smoothly as this transition occurs. The abrupt changes to  the
$\varpi$ coordinate  occur because very small perturbations to 
very nearly circular orbits
produce large changes to this angle.
}

\subsection{Multiple Planets}
Up to now we have considered a single planet. However, it is a simple matter
to generalize the above formalism so that it applies 
to a system of  two planets. Here we follow closely the discussions
in \cite{PapaloizouResonances} and 
\cite{PapaloizouSzuszkiewicz2005}. 
Excluding  stochastic forcing for the time being,  we start from the 
Hamiltonian formalism describing their mutual
interactions 
  using Jacobi coordinates \citep[see][]{Sinclair1975}.
 In  this formalism the radius vector
 ${\bf r}_2,$ of the inner planet of reduced  mass $m_2$
is measured from  $M$
and that of the  outer planet, ${\bf r }_1,$
of reduced  mass $m_1$ is referred
to the  centre of mass of  $M$ and  $m_2.$
Thus from now on we consistently adopt a subscripts  $1$ and $2$
for coordinates related to the outer and inner planets respectively.

The required Hamiltonian  correct to second order in the planetary masses  is given by 
\begin{eqnarray} H & = &  {1\over 2} ( m_1 | \dot {\bf r}_1|^2 +m_2| \dot {\bf r}_2|^2)
- {\G M_{1}m_1\over  | {\bf r}_1|} - {\G M_{2}m_2\over  | {\bf r}_2|} \nonumber \\
& - &{\G m_{1}m_2\over  | {\bf r}_{12}|}
 +  {\G m_{1}m_2 {\bf r}_1\cdot {\bf r}_2
\over  | {\bf r}_{1}|^3}.
\end{eqnarray}
Here $M_{1}=M+m_1$, $M_{2}= M + m_2$ and
${\bf r}_{12}= {\bf r}_{2}- {\bf r}_{1}$.
The Hamiltonian can be expressed in terms of 
 $E_i,G_i,\varpi_i, \lambda_i,  i=1,2$ 
 and the time
$t.$ 
The energies are  given by  $E_i = -\G m_iM_{i}/(2a_i),$
and the angular momenta $G_i = m_i\sqrt{\G M_{i}a_i(1-e_i^2)}$  with 
 $a_i$ and $e_i$ denoting the semi-major axes and eccentricities
respectively.    The mean motions are 
$n_i = \sqrt{\G M_{i}/a_i^3}.$

\noindent The Hamiltonian may quite generally
 be expanded in a Fourier series
involving linear combinations of the three angular differences
$\lambda_i - \varpi_i, i=1,2$ and $\varpi_1 -\varpi_2$
\citep[eg.][]{BrouwerClemence1961}.

Near a first order  $p+1 : p $ resonance, we expect that both
$\phi_1 = (p+1)\lambda_1-p\lambda_2-\varpi_2, $ and
$\phi_2 = (p+1)\lambda_1-p\lambda_2-\varpi_1,$
will be slowly varying.
 Following  standard practice \citep[see eg.][]{PapaloizouResonances, PapaloizouSzuszkiewicz2005} 
only terms in the Fourier expansion involving  linear 
combinations of $\phi_1$ and $\phi_2$
as argument are  retained because 
only  these are expected to lead to large long term perturbations.

\noindent The resulting  Hamiltonian   may then be written
in the general form $H=E_1+E_2+ H _{12},$ where
\be H _{12}= -\frac{\G m_1m_2}{a_1}\sum C_{k,l}\left( \frac{a_1}{a_2}, e_1,e_2
\right) \cos (l\phi_1 +k\phi_2), \label{Hamil} \ee
where in the above and similar summations below, the sum   ranges
over all positive
and negative integers $(k,l)$  and  the dimensionless  coefficients  $C_{k,l}$
depend on $e_1,e_2$ and  the ratio $a_1/a_2$ only.
We also  
replace $M_{i}$ by $M.$

\subsubsection{Equations of Motion}
The equations of motion for each planet can now be easily derived that take into account
the contributions due to their 
to mutual interactions \citep[see][]{PapaloizouResonances, PapaloizouSzuszkiewicz2005} and contributions from (\ref{feqnmo1}) - (\ref{feqnmo4}). 
The latter terms arising from external forcing are indicated with a subscript $F$. 
We thus obtain to  lowest order in the  perturbing masses.
\begin{eqnarray}
{dn_1\over dt} & = & {3(p+1)n_1^2 m_2 \over M  } \sum C_{k,l}(k+l)\sin (l\phi_1 +k\phi_2)\nonumber \\
    && + \left(dn_1\over dt\right)_F\label{first}\\
{dn_2\over dt} &  = & -{3pn_2^2 m_1 a_2\over M a_1 }\sum C_{k,l}(k+l)\sin (l\phi_1 +k\phi_2)\nonumber \\
    && + \left(dn_2\over dt\right)_F\label{first1}\\
{de_1\over dt} &=&  - {m_2 n_1 \sqrt{1-e_1^2} \over  e_1 M } \cdot  \sum C_{k,l}\sin (l\phi_1 +k\phi_2) \label{first2}  \\
    && \cdot
    \left[k-(p+1)(k+l)\left(1-\sqrt{1-e_1^2}\right)\right] + \left(de_1\over dt\right)_F\nonumber\\ 
{de_2\over dt}  &=& - {m_1 a_2  n_2 \sqrt{1-e_2^2} \over a_1 e_2 M }\cdot \sum C_{k,l}\sin (l\phi_1 +k\phi_2) \nonumber \\
    && \cdot 
    \left[l+p(k+l)\left(1-\sqrt{1-e_2^2}\right)\right]  + \left(de_2\over dt\right)_F   \label{first3}\\
{d\phi_2 \over dt} & =&  (p+1)n_1- pn_2 -
    \sum  (D_{k,l} +E_{k,l})\cos(l\phi_1 +k\phi_2)\nonumber\\
     && +  \left(d\phi_2\over dt\right)_F  \label{last2} \\
{d\phi_1 \over dt} &=&    (p+1)n_1- pn_2 -
    \sum ( D_{k,l} + F_{k,l} )\cos(l\phi_1 +k\phi_2)\nonumber\\
    && + \left(d\phi_1\over dt\right)_F  \label{last} .
 \end{eqnarray}
Here
 \begin{eqnarray} 
D_{k,l} &=& {2(p+1)n_1a_1^2m_2\over M}{\partial \over \partial a_1}\left(  C_{k,l}/a_1 \right)\nonumber \\
   &&- {2pn_2a_2^2m_1 \over M}{\partial \over \partial a_2}\left(  C_{k,l}/a_1 \right),\\
E_{k,l} & = & {n_1m_2 \left((p+1)(1 -e_1^2)-p\sqrt{1-e_1^2}\right) \over e_1 M}{\partial  C_{k,l}\over \partial e_1}\nonumber \\
   && + {pn_2a_2m_1\left(\sqrt{1-e_2^2}-1+e_2^2\right) \over a_1 e_2 M}{\partial  C_{k,l}
\over \partial e_2} 
\end{eqnarray}
and
\begin{eqnarray}
 F_{k,l} & = & {(p+1)n_1m_2 \left(1 -e_1^2-\sqrt{1-e_1^2}\right) \over e_1 M}{\partial  C_{k,l}\over \partial e_1} \nonumber \\
    && +  {n_2a_2m_1\left((p+1)\sqrt{1-e_2^2}-p(1-e_2^2)\right) \over a_1 e_2 M}{\partial  C_{k,l} \over \partial e_2}.
\end{eqnarray}
Note that $\phi_2 - \phi_1 = \varpi_2 - \varpi_1\equiv \Delta \varpi $
is the  angle between the two  apsidal lines of the two planets.
We also comment that, up to now, 
 we have not assumed that the eccentricities are small and that, in additional
 to stochastic contributions,  the external forcing terms may in general
contain contributions from very slowly varying disk tides
but we shall not consider these further in this article.

\subsubsection{Modes of Libration}\label{FSmodes}
We first consider two planets in resonance with no external forces acting
in order to identify the possible libration modes. We then consider the effects
of the addition of  external stochastic forcing.
In the absence of external forces equations (\ref{first}) - (\ref{last}) can have a solution
for which  $a_i$ and $e_i$ are constants and the angles $\phi_1$ and $\phi_2$ are zero.
In general other values for the angles may be possible but such cases do not occur
for the numerical examples presented below. When the angles are zero
equations (\ref{last2}) and (\ref{last}) provide a relationship between $e_1$ and $e_2$ \citep[see eg.][]{PapaloizouResonances}.

 We go on to consider small amplitude oscillations or
librations of the angles about their above equilibrium  state. Because two planets are involved there are
two modes of oscillation which we find convenient to separate and describe as
fast and slow modes.  Assuming the planets have comparable masses, the
fast mode has libration frequency $\propto \sqrt{m}$
and the slow mode has libration frequency  $\propto m.$
These modes clearly separate as the planet masses are decreased while
maintaining  fixed eccentricities.

\subsubsection{Fast Mode}
To obtain the fast mode we linearize (\ref{first})~-~(\ref{last})
and neglect second order terms in the planet masses. This is equivalent
to neglecting the variation of 
$D_{k,l}, E_{k,l}$ and $ F_{k,l}$ in  equations (\ref{last2}) and (\ref{last})
which then require that $\phi_1 = \phi_2$  very nearly for this mode.
\changed{Noting that for linear modes of the type considered here
and in the next section, equations (\ref{first})-(\ref{first3}) imply that $\dot n_i$ and $\dot e_i$ are proportional to linear combinations of the librating angles,} 
differentiation of either of equations (\ref{last2}) 
or (\ref{last}) with respect to time then gives for small amplitude oscillations
 \begin{eqnarray}
 {d^2\phi_i \over dt^2} 
+\omega_{lf}^2 \phi_i &=& 0, \ \ \ ( i= 1,2 ), 
\label{fast} \end{eqnarray}
where
\begin{equation}
\omega_{lf}^2= - {3p^2n_2^2 m_2 \over M }
\left(1+{ a_2 m_1\over a_1m_2}\right)\cdot \sum C_{k,l}(k+l)^2\label{fastp}
\end{equation}
and we have used the resonance condition that $(p+1)n_1 = p_2 n_2$
which is satisfied to within a correction of order $\sqrt{m_1/M}.$
Note that for this mode the fact that $\phi_1 = \phi_2$ very nearly, implies that
$\varpi_2 -\varpi_1$ is small. Thus  that quantity does not participate in the oscillation.

\subsubsection{Slow Mode}
In this case we look for low frequency librations with
frequency $\propto m_1.$ Equations (\ref{last2}) and (\ref{last})
imply that, for such oscillations, to within a small relative error of order $m_1/M,$ 
$(p+1)n_1 = pn_2$ throughout. 

Equations (\ref{first}) and (\ref{first1}) then imply
that  the two angles are related by $\phi_2=\beta \phi_1,$ 

\noindent where 
$\beta  = - \sum C_{k,l}(k+l)l)/(\sum C_{k,l}(k+l)k).$
Subtracting equation (\ref{last})  from equation (\ref{last2}), 
differentiating with respect
to time  and using this condition results in an equation

\noindent for $\zeta = \phi_ 2- \phi_1 = \varpi_2 -\varpi_1 = \Delta \varpi$ 
 \begin{eqnarray}
 {d^2\zeta \over dt^2} 
+\omega_{ls}^2 \zeta &=& 0,
\label{slow} \end{eqnarray}
where
\begin{equation}
\omega_{ls}^2= \alpha_1 {n_1m_2\sqrt{1 -e_1^2} 
\over  e_1 M (1-\beta)}{\partial W \over \partial e_1} + \alpha_2  {n_2a_2m_1\sqrt{1 -e_2^2} 
\over a_1 e_2 M (1-\beta)}{\partial W \over \partial e_2}\label{slowp}
\end{equation}
with
\begin{eqnarray}
\alpha_1 &= & 
\sum C_{k,l}\left[k-(p+1)(k+l)\left(1-\sqrt{1-e_1^2}\right)\right](k\beta+l),\nonumber\\
\alpha_2 & =&  
\sum C_{k,l}\left[l+p(k+l)\left(1-\sqrt{1-e_2^2}\right)\right](k\beta+l)\nonumber
\end{eqnarray}
and
\begin{eqnarray}
 W & = & \left( {n_1a_2m_2\sqrt{1 -e_2^2} \over a_1 e_2 M}{\partial \over \partial e_2}
-{n_2m_1 \sqrt{1 -e_1^2} \over e_1 M}{\partial \over \partial e_1} \right)\nonumber\\
&& \cdot \sum C_{k,l}. \nonumber
 \end{eqnarray}
Although the expressions for the mode frequencies are complicated,
the fast frequency scales as the square root of the planet mass
and the slow frequency scales as the planet mass independent of the magnitude of the eccentricities
while both scale as the characteristic  mean motion of the system.
Furthermore although we have considered small amplitude librations
and accordingly obtained harmonic oscillator equations, the treatment can be extended
to consider finite amplitude oscilations and generalized pendulum equations
as long as the two mode frequencies can be separated.
However, we shall not consider this aspect further here.

\subsubsection{Librations with External Forcing}
When external forcing is included source terms appear
on the right hand sides of equations (\ref{fast}) and (\ref{slow}).
We shall assume that the forcing terms are small so that terms involving
products of these and both the \changed{libration amplitudes and the}  
 planet masses may be neglected.
Then in the case of the slow mode, repeating the derivation given above
including the forcing terms, we find that (\ref{slow}) becomes
\begin{eqnarray}
 {d^2\zeta \over dt^2} 
+\omega_{ls}^2 \zeta &=& {d\over dt}\left(\dot \varpi_{2F} - \dot \varpi_{1F} \right).
\label{slowf} \end{eqnarray}
The quantities $\dot\varpi_{iF}$ are readily obtained for each planet from  (\ref{feqnmo3}).
From this we see that for small eccentricities, $\dot\varpi_{iF} \propto 1/e_i,$
indicating  large  effects when $e_i$ is small. \changed{As already discussed
in section \ref{AdF} this feature arises from a coordinate singularity
rather than physically significant changes to the system.}

A similar description may be found for the fast mode. In this case, neglecting terms
of the order of the square of the planet masses or higher,  one may
use  equations (\ref{last2}) 
and  (\ref{last}) to obtain an equation for $Q \equiv \phi_1 $ in the form
\begin{eqnarray}
 {d^2 Q \over dt^2} 
+\omega_{lf}^2 Q&=&{d\over dt}\left[
 \dot \phi_{1F} \right ]\nonumber \\
&&+(p+1)\dot n_{1F}-p\dot n_{2F}.
\label{fastf} \end{eqnarray}

\changed{ If equations for quantities that are regular functions of 
the Cartesian like coordinates $(h_i=e_i\sin\varpi_i \mbox{ and } k_i=e_i\cos\varpi_i,\, i=1,2)$ are found,  the source terms arising from the external forcing are regular as $e_i \rightarrow 0$. To illustrate this we consider the pair}
\begin{eqnarray}
y&\equiv& e_1e_2\sin\zeta= h_2k_1-h_1k_2\quad \mbox{ and} \nonumber\\
z&\equiv& e_1e_2\cos\zeta= k_1k_2-h_1h_2. \nonumber
\end{eqnarray}
 \changed{It is readily seen that their time derivatives are given by }
 \begin{eqnarray}
{\dot y}&=&{d(e_1e_2)\over dt}\sin\zeta + e_1e_2\cos\zeta\left[\left.{d\zeta\over dt}\right|_U 
 + \left(\dot \varpi_{2F} - \dot \varpi_{1F} \right)\right]\label{revise0}\\
{\dot z}&= & \cos\zeta\left[\left.{d(e_1e_2)\over dt}\right|_U 
 +  e_2\dot e_{1F} + e_1\dot e_{2F}\right]- e_1e_2\sin\zeta{d\zeta\over dt}
 \label{revise}\end{eqnarray}
\changed{Here $|_U $ denotes evaluation without external forcing.
There is now no divergence of the external forcing terms as 
$e_i \rightarrow 0.$
If as above small amplitude slow mode  librations of $\zeta$ about zero are considered,
given that without external forcing $\dot e_i \propto \zeta,$ the first term on the right hand side
of (\ref{revise0}) is either quadratic in $\zeta,$ 
or proportional to the product of the exernal forcing  and libration amplitudes
and so,  as above  may be neglected. The same applies to last term on the right hand side
of (\ref{revise}). In addition $\cos\zeta$ may be replaced by   
unity. On taking the time derivatives of (\ref{revise0}) and (\ref{revise}),
we obtain under the same approximation  scheme} 

\begin{eqnarray}
 {\ddot y}& =&      
  e_1e_2\left.{d^2\zeta\over dt^2}\right|_U
+{d \over dt}\left[ e_1e_2\left( \dot \varpi_{2F} - \dot \varpi_{1F}\right)\right]
 \label {heq0}\\
{\ddot z}& =&      
  e_2\left.{d^2 e_1\over dt^2}\right|_U + e_1\left.{d^2 e_2\over dt^2}\right|_U
+{d \over dt}\left[ e_1 \dot e_{2F} 
+ e_2\dot e_{1F} \right]
. \label {heq}
\end{eqnarray}
\changed{Making use
 of (\ref{slow}) and its counterparts for ${\ddot e_1}$ and ${\ddot e_2}$ for the unforced motion we obtain}
\begin{eqnarray}
 {\ddot y}& = &      
   -\omega_{ls}^2 y
 +{d \over dt}\left[ e_1e_2\left( \dot \varpi_{2F} - \dot \varpi_{1F}\right)\right]\label{heq10}\\
{\ddot {\delta{z}}}& = &      
   -\omega_{ls}^2 \delta  z
 +{d \over dt}\left[ e_1 \dot e_{2F} + e_2\dot e_{1F}\right]\label{heq1}
, \end{eqnarray}
\changed{where $\delta z= z-z_0,$ $z_0$ being the value of  $e_1e_2$ at the centre of the oscillation in $z.$  
These may be used as an alternative to (\ref{slowf}),  and are without potentially
singular forcing terms. We note that for the purpose of this paper, both descriptions
  provide  exactly the same physical 
content.}

Equations (\ref{fastf}) and (\ref{slowf})\changed{ (or (\ref{heq10}) together with (\ref{heq1}))}
  form a pair of equations
for the stochastically forced fast and slow modes respectively.
We comment that this mode separation is not precise.
However, it can be made so by choosing  appropriate linear
combinations of the above modes. 
Numerical results confirm that $Q$ predominantly
manifests the fast mode  and $\zeta$ \changed{or $y$} the slow mode, so we do not
expect such a change of basis to significantly affect conclusions.

We  further  comment that because $\dot\phi_{1F}$  contains $\dot\varpi_{2F}$
but not $\dot\varpi_{1F},$ for small eccentricities there are only potential forcing terms $\propto$~$1/e_2$
that occur when forcing is applied to the inner planet.
As this  planet has the larger ecentricity for the situations we consider, small eccentricities
are not found to play any significant role in this case.

Each mode responds as  a forced
oscillator. We suppose the forcing contains a stochastic component
which tends to excite the respective oscillator and ultimately convert
libration into circulation. But we stress that the above formulation
as well as developments below assume small librations, so we may only assess
the initial growth of oscillation amplitude. However, inferences based on the 
structure of the non linear governing equations and an extrapolation 
of the linear results enable successful comparison with numerical results.

\subsection{Stochastic Forces}\label{StochF}

We assume that turbulence 
causes the external force per unit mass $(F_r, F_{\theta})$  acting on each planet to be stochastic.
For simplicity we shall
adopt the simplest possible model. Regarding the components of the force, 
per unit  mass expressed in cylindrical
coordinates to be a  function of time $t,$ we assume any one of these
satisfies the relation $F_i(t)F_i(t')= \left< F_i^2 \right> g(|t-t'|)  $
 where the autocorrelation function $g(x)$ is such that
 $\int^{\infty}_0 g(x) dx = \tau_c,$
where $\tau_c$ 
is the correlation time and $\sqrt{\left< F_i^2 \right>}$ is the root mean square
value of the $i$ component. It should be noted that an ensemble
average is implied on the left hand side. Again for simplicity we assume that different components 
acting on the same planet as well as components acting on different planets are uncorrelated. We note that in general the root mean square values as well as 
$\tau_c$  may depend
on $t,$ but we shall not take this into account here and simply assume that these
quantities are constant \changed{and the same for each force component}. 

We note the stochastic forces make quantities they act on undergo a random walk. Thus 
if for example  $\dot A = F_i$ for some quantity $A$ (note that constants or slowly varying
quantities originally multiplying $F_i$ may be absorbed by a redefinition
of $A$ and so do not materially affect the discussion given below), the square of the change of $A$ occuring
after a time interval $t$ is given by
\begin{eqnarray}
(\Delta A)^2 &=& \int^t_0 \int^t_0 F_i(t')F_i(t'')dt'dt'' \nonumber \\
&=& \int^t_0 \int^t_0  \left< F_i^2 \right>g(|t'-t''|) dt'dt'' \rightarrow  D_i t,
\label{stoch}
\end{eqnarray}
Here we take the limit where $t/\tau_c $  is very large corresponding
to an integration time  of very many correlation times and 

\noindent $D_i = 2\left< F_i^2 \right>\tau_c$ is the diffusion coefficient. 

When the evolution of a stochastically forced
planetary orbit is considered, it is more appropriate to consider a model  governing equation for $A$
of the generic form
 \begin{equation} \dot A = F_i \sin(nt),\label{generic} \end {equation} 
  where we recall that $2\pi/n$ is the orbital period (but note
 that a different value could equally well be considered).
 We note  in passing that, by shifting the origin of time,
  an arbitrary phase may be added to the argument of the $\sin$
 without changing the results given below. One readily finds that 
 equation (\ref{stoch}) is replaced by  
\begin{eqnarray}
(\Delta A)^2&=& \int^t_0 \int^t_0 F_i(t')F_i(t'')\sin(n t') \sin(n t'') dt'dt''\nonumber \\
&=&\int^t_0  \int^t_0\left< F_i^2 \right>g(|t'-t''|) \sin(n t')\sin(n t'')dt' dt'' \nonumber  \\
&\rightarrow& {\gamma Dt\over 2},
\label{stochreduced}
\end{eqnarray}
where $\gamma(n) = \int^{\infty}_0 g(x)\cos(n x) dx .$

\noindent Note that when  $n\tau_c\ll 1,$ corresponding
to the correlation time being much less than the orbital
period,  $\gamma \rightarrow 1.$
For larger $\tau_c,$ $\gamma <1$ gives a reduction factor
for the amount of stochastic diffusion. For example if we adopt
an exponential form for the autocorrelation function such that
\begin{eqnarray*}
	g(|t'-t''|) = \exp \left(- \frac{|t'-t''|}{\tau_c}\right),
\end{eqnarray*}
we find 
\begin{eqnarray}
 \gamma = \frac 1{1+n^2 \tau_c^2}\label{eq:reduction}
\end{eqnarray}
and for the purposes of comparison
with numerical work 
we shall use this from now on.

\subsubsection{Stochastic Forcing of an Isolated Planet}
We begin by considering the effect of stochastic forcing on a single isolated planet.
In this case we may obtain  a statistical estimate for the 
characteristic growth of the orbital parameters 
as a function of time by integrating  equations (\ref{feqnmo2}), (\ref{feqnmo3}), 
 \changed{(\ref{feqnmo5}),  (\ref{feqnmo7}) and  (\ref{feqnmo8})}
with respect to time directly. 
We may then apply the formalism leading to 
the results  expressed  in generic form
through equations (\ref{stoch})~-~(\ref{eq:reduction})
 to  obtain estimates for the stochastic diffusion
 of the orbital elements  in the limit of small eccentricity in the form
\begin{eqnarray}
 (\Delta a)^2 &=& 4 \frac {  D t }{ n^2 }\label{eq:growtha}\\
 (\Delta e )^{2} &=& 2.5 \frac {\gamma D t }{ n^2 a^2 }  \label{eq:growthe}\\
 (\Delta \varpi)^2 &=& \frac{2.5}{e^2} \frac{\gamma Dt}{n^2 a^2}  \\ \label{eq:growthw}
 (\Delta h )^{2} &=& 2.5 \frac {\gamma D t }{ n^2 a^2 }  \label{eq:growthh}\\
 (\Delta k )^{2} &=& 2.5 \frac {\gamma D t }{ n^2 a^2 }  \label{eq:growthk}.
\end{eqnarray}

\changed{ We note again that the $1/e^2$ dependence of $(\Delta \varpi)^2$
which arises from the coordinate singularity discussed in section \ref{AdF}
does not appear   for $(\Delta h )^{2}$ and $ (\Delta k )^{2}.$
Note that the definition of $(h,k)$ imply consistently with the above  that}
\begin{eqnarray}
 (\Delta h )^{2} &=& (\Delta e )^{2}\sin^2\varpi+  e^{2} (\Delta \varpi)^2\cos^2\varpi   \label{eq:growthh1}\\
 (\Delta k )^{2} &=& (\Delta e )^{2}\cos^2\varpi+  e^{2} (\Delta \varpi)^2\sin^2\varpi  \label{eq:growthk1}.
\end{eqnarray}

\subsubsection{Stochastic Variation of the 
Resonant Angles in the Two Planet Case} \label{STF2}
 
 We now consider the effects of stochastic forcing on the resonant angles. \changed{The expressions are more complicated than in the previous section because there are more variables involved. However, we basically follow the formalism outlined in section \ref{StochF}.}
 
 While the libration amplitude is small enough
 for linearization to be reasonable, the evolution is described by
 equations  (\ref{fastf}) and (\ref{slowf}) (\changed{or (\ref{heq10}) together with (\ref{heq1}))}.
These may be solved by the method
 of variation of parameters. Assuming the amplitude is zero at $t=0,$
 the solution of \changed{equation (\ref{heq10})} is given by 
 \begin{eqnarray}
 y &=& \sin(\omega_{ls}t)\int^t_0{S_y\cos(\omega_{ls}t)\over \omega_{ls}}dt \nonumber\\
 && - \cos(\omega_{ls}t)\int^t_0{S_y\sin(\omega_{ls}t)\over \omega_{ls}}dt,
 \label{zetasol}
 \end{eqnarray}
\changed{where $S_y= {d}(e_1e_2(\dot\varpi_{2F}- \dot \varpi_{1F}))/dt.$}
 There   are  corresponding expressions that can be obtained \changed{from 
 equations (\ref{heq1}) and  (\ref{fastf}) 
 for $\delta z$ and} $Q$ respectively.

 Equation (\ref{zetasol}) may be regarded as describing
 a harmonic oscillator whose amplitude varies in time such that
 the square of the amplitude after a time interval
 $t$ is given by
 \begin{equation}
 (\Delta y )^2~=~\left(\int^t_0{S_y\sin(\omega_{ls}t)\over \omega_{ls}}dt\right)^2
 ~+~\left(\int^t_0{S_y\cos(\omega_{ls}t)\over \omega_{ls}}dt\right)^2.
 \label{Dzeta} \end{equation}
The  corresponding expression 
  from  \changed{equation (\ref{heq1}) is} 
 \begin{equation}
 (\Delta \delta z )^2~=~\left(\int^t_0{S_z\sin(\omega_{ls}t)\over \omega_{ls}}dt\right)^2
 ~+~\left(\int^t_0{S_z\cos(\omega_{ls}t)\over \omega_{ls}}dt\right)^2,
 \end{equation}  
  \changed{where $S_z = {d}(e_1\dot e_{2F}+e_1\dot e_{2F})/dt$. And finally the equation obtained from} equation (\ref{fastf}) is
 \begin{equation}
 (\Delta Q)^2 = \left(\int^t_0{S_Q\sin(\omega_{lf}t)\over \omega_{lf}}dt\right)^2
 +\left(\int^t_0{S_Q\cos(\omega_{lf}t)\over \omega_{lf}}dt\right)^2,
 \label{DQ} \end{equation}
where  $S_Q= d(\dot \phi_{1F})/dt
+(p+1)\dot n_{1F} - p\dot n_{2F}.$

We now evaluate the expectation values of these using the
formalism of section \ref{StochF}.
For simplicity and as considered numerically later we shall specialize to the case
when stochastic forces act only on the outer planet (but see section \ref{sec:discussion} below).   Taking 
equation (\ref{Dzeta}), we  perform an integration by parts  neglecting
the end point contribution as these are associated with subdominant contributions
 increasing less rapidly than $t$ for large $t,$  to obtain
 \begin{eqnarray}
 (\Delta y )^2 &=& (\Delta (e_1e_2)\sin\zeta )^2 \nonumber\\  
 &=& \left|\int^t_0{e_1e_2\dot \varpi_{1F}\sin(\omega_{ls}t)}dt\right|^2 \nonumber \\
 && + \left|\int^t_0{e_1e_2\dot\varpi_{1F}\cos(\omega_{ls}t)}dt\right|^2.
 \label{Dzetaf} \end{eqnarray}
 \changed{We then find, working
 in the limit of a small eccentricity $e_1$ (but not necessesarily $e_2$), from the corresponding equation for $\delta z$
 that $(\Delta \delta z )^2 = (\Delta y )^2 .$}
 
 In dealing with equation (\ref{DQ}) we neglect 
$\dot \phi_{1F}$ in $S$ because after integration by parts 
this leads to a  contribution on the order
$\omega_{lf}/n$ smaller than that derived from $\dot n_{1F}.$ 
Thus we simply
obtain
 \begin{eqnarray}
 {(\Delta Q)^2\over (p+1)^2}
 &=& \left(\int^t_0{\dot n_{1F}\sin(\omega_{lf}t)\over \omega_{lf}}dt\right)^2 \nonumber\\
 &&+\left(\int^t_0{\dot n_{1F}\cos(\omega_{lf}t)\over \omega_{lf}}dt\right)^2.
 \end{eqnarray}
We now  follow the procedures outlined in section \ref{StochF} obtaining
\begin{eqnarray}
 (\Delta e_1e_2\sin\zeta )^2 &=& 
 2.5 {De_2^2 \gamma_s t\over  2a_1^2 n_1^2 }\label{eq:growths}\quad \quad \text{and} \\
  {(\Delta Q)^2\over (p+1)^2}
 &=& {9 D \gamma_ft\over a_1^2\omega_{lf}^2} 
 \label{eq:growthf}
 \end{eqnarray}
where 
\begin{eqnarray}
\gamma_f &=& \frac1{1+\omega_{lf}^2\tau_c^2} \quad \quad \text{and}\\
\gamma_s &=& \frac1{1+(n_1+\omega_{ls})^2\tau_c^2} +\frac1{1+(n_1-\omega_{ls})^2\tau_c^2 }.
\end{eqnarray}

\subsubsection{Growth of Libration Amplitudes}\label{GRLIB}
Equations (\ref{eq:growths}) and (\ref{eq:growthf}) express the expected
growth of the resonant angle libration amplitudes as a function of time.
We  remark that these expressions  can be simply related to those
obtained for a single planet. 
Thus equations (\ref{eq:growtha}) and (\ref{eq:growthf}) applied 
to the outer planet imply that \newline
\begin{equation}
(\Delta Q)^2
/(\Delta a_1)^2=  
9(p+1)^2 n_1^2\gamma_f /(4 a_1^2\omega_{lf}^2). \nonumber
\end{equation}
As we are interested in the case $p=1,$  the width of the libration zone
is $\sim a_1\omega_{lf}/n_1,$ we see that the time for $(\Delta Q)^2$ to reach unity is comparable
for the semi-major axis to diffuse through the libration zone.

Similarly, \changed{for small amplitude librations about fixed  $e_2$  equation (\ref{eq:growths}) 
gives almost the same result   for $(\Delta e_1\sin\zeta )^2$  
or $e_2^{-2}(\Delta \delta z)^2$ as that  
obtained for $(\Delta h )^2$ or $(\Delta k)^2$ from  equation (\ref{eq:growthw})}
applied to the outer planet. 
This  indicates that $\zeta$, being the angle between the apsidal 
lines of the two planets, diffuses  in the same way as for an isolated outer planet subject to 
stochastic forces. Thus, in the small amplitude regime,
the way this diffusion occurs would appear to be
essentially independent of whether the planets are 
in resonance (but see below).

 \changed{An important   consequence of equation (\ref{eq:growths}) is the behaviour 
of $\zeta$ for small $e_1.$ 
The latter quantity was assumed constant in the analysis.
  As implied by the discussion of section \ref{AdF}
abrupt changes to $\zeta$ are expected when $(y,z)$ passes through the
origin. Then even an initially small amplitude libration is converted to circulation.
Thus if $e_1$ is small then equation (\ref{eq:growths}) indicates that a
  time $  t \sim
  4a_1^2 n_1^2  e_1^2 / (5 D \gamma_s \zeta^2 )$, is required to convert libration to circulation.
This can be small if $e_1$ is small. Even if $e_1$ is not small initially, 
it is important to note that}  it
 also undergoes stochastic diffusion 
(see equation (\ref{eq:growthe})) as well as
oscillations through its participation in libration. 
Should $e_1^2$
attain {\it very small} values through this process, 
then from (\ref{eq:growths}) we expect the onset of a rapid 
evolution of $\zeta.$ 
Accordingly the attainment of circulation for this angle,  should be related
to  the diffusion of $e_1^2$ allowing very small values of that quantity to be attained, rather than the direct excitation of libration amplitude. 
This is particularly the case
when $e_1^2$ starts from relatively small values.

In fact, application of (\ref{eq:growths}) and (\ref{eq:growthf})
to the numerical examples discussed below \changed{adopting the initial orbital elements}
 indicates that the diffusion
of $\zeta$ is significantly smaller than $Q$ unless  $e_1$ starts out
with a  very small value. This would suggest that $Q$ reaches circulation 
before $\zeta.$ However, this  neglects the coupling between the angles
that occurs once the libration amplitudes become significant.
It is readily seen that it is not expected that $Q$ could circulate
while $\zeta$ remains librating as it was initially.
 One expects to recover standard
secular dynamics for $\zeta$
  from the governing equations (\ref{first})~-~(\ref{last2}), 
when these are averaged  over an assumed  $Q=\phi_1$ circulating
with constant $\dot Q.$ As a libration of the initial
form would not occur under those conditions, we expect, and find, 
large excursions or 
increases in the libration amplitude of $\zeta$ to be correlated with
increases in the libration amplitude of $Q.$ 
This in turn increases the oscillation amplitude of the eccentricity $e_1$,
allowing it to  approach zero. The consequent rapid \changed{evolution} of $\zeta$
then enables it to pass to circulation.
Thus the breaking of resonance is ultimately found to be  controlled by the 
excitation of large amplitude librations for $Q =\phi_1,$
which induce $\zeta$ to pass to circulation somewhat before
$Q$ itself.

\section{Numerical Simulations}

\begin{table}
\centering                        
\begin{tabular}{ r | c c c c c }       
\hline                 
\hline                 
Planet & $m$ ($\rm{M}_J$) & $a$ (AU) & $P$ (days)  &$e$ & $ \zeta$  \\   
\hline   
GJ876  c & 0.790 & 0.131 &30.46& 0.263 & $10\degree $ \\ 
       b & 2.530 & 0.208 &60.83& 0.031 &  \\
\hline    
GJ876 LM  c & 0.13 & 0.131 &30.46& 0.263 & $10\degree $ \\ 
       b & 0.42 & 0.208 &60.83& 0.031 &  \\
\hline    
GJ876 SE  c & 0.013 & 0.131 &30.46& 0.263 & $10\degree $ \\ 
       b & 0.042 & 0.208 &60.83& 0.031 &  \\
\hline    
GJ876 E  c & 0.0013 & 0.0131 &30.46& 0.263 & $10\degree $ \\ 
       b & 0.0042 & 0.208 &60.83& 0.031 &  \\
\hline    
GJ876 LM HE c & 0.13 & 0.131 &30.46& 0.41 & $10\degree $ \\ 
       b & 0.42 & 0.208 &60.83& 0.09 &  \\
\hline    
GJ876 SE HE c & 0.013 & 0.131 &30.46& 0.41 & $10\degree $ \\ 
       b & 0.042 & 0.208 &60.83& 0.09 &  \\
\hline    
HD128311 b & 1.56 & 1.109 & 476.8 & 0.38 & $58 \degree$ \\
         c & 3.08 & 1.735 & 933.1 & 0.21 & \\
\hline                               
\end{tabular}
\caption{Parameters of the model systems considered. 
The first has orbital elements taken from the three planet fit to GJ876 with orbital inclination
to the plane of the sky, 
 $i~=~50\degree$ \citep{Rivera2005}\label{table:initorb}.
   The table entries labelled as  LM, SE and E have the same parameters as the first
  entry but the planet masses are scaled 
  down by  constant factors of 6, 60, and 600.
  Systems with the  added label  HE have 
  larger orbital eccentricities.
 The final entry  is a system with the observed elements of
 HD128311 \citep{Vogt2005}. }
     
\end{table}

We have performed numerical simulations of one and two planet systems that allow for the  incorporation
 of additional stochastic forces with the properties described above. These in turn provide
a simple prescription for estimating the effects of stochastic gravitational forces produced by density
fluctuations associated with disk turbulence.
Results have been obtained using both a  fifth order  Runge Kutta method and  
the  Bulirsch Stoer method \citep{StoerBulirsch02,nr} 
with fixed as well as adaptive timesteps. We have checked that
 results are  converged and thus do not actually depend on the  integrator used.

First we  discuss the expected scaling of the stochastic forces  with the physical parameters of the disk
and their implementation in  the n-body integrations. In order to clarify the physical mechanisms
involved and  to check the analytic predictions
for stochastic diffusion given by equations (\ref{eq:growtha}) - (\ref{eq:growthw})  we   
consider simulations of  a single planet   undergoing stochastic forcing first. 
We then move on to consider    two planet commensurable systems 
with and without stochastic forcing.  We focus on the way a  2:1 commensurability,
corresponding to $p=1$   is disrupted,
highlighting the various evolutionary stages  a system goes through as it evolves
from a state with a strong commensurability affecting the interaction dynamics, to one
where the commensurability is completely disrupted and in some cases
a strong scattering occurs. 
We consider  a range of different planet masses and eccentricities (see table \ref{table:initorb}).

\subsection{Stochastic Forces}\label{sec:scaling}
In order to mimic the effects of 
turbulence, for example produced by the  MRI, 
it is necessary to calibrate these forces with reference to MHD simulations. 
As described above, the basic  parameters characterising
the prescription for stochastic forcing
that we have implemented
are the root mean square value of the force components 
per unit mass (in cylindrical coordinates)  $\sqrt{\left< F_i^2 \right>}$ 
and the auto correlation time $\tau_c$. 

 From our analytic considerations, we concluded 
the stochastic forces make the orbital parameters undergo a random walk 
that  is dependent on the force  model primarily 
through the diffusion  coefficient $D = 2\left< F_i^2 \right>\tau_c .$

For planets under the gravitational influence 
of a protoplanetary disk, the  natural scale for the force per unit mass components,~$F_i$, is 
$F_0(r) = \pi \G\Sigma(r)/2$, where $\Sigma$ is the characteristic disk surface density \citep[see eg.][]{papaloizouterquem06}.
We comment that $F_0(r)$ is  the gravitational force per unit mass due to a small circular disk patch
of radius $r_{\Sigma}$ at a distance $\sqrt{2} \; r_{\Sigma}$ from its centre assuming
that all its mass is concentrated there. The result  is independent of $r_{\Sigma}.$ 
The natural  correlation time $\tau_c$ is the  inverse of the orbital angular frequency 
  $\tau_{c,0} = \Omega^{-1}$.  To set the natural scale for $D,$ we adopt a
 minimum mass solar nebula model  \citep[MMSN, see][]{Weidenschilling1977} with   
\begin{eqnarray}
\Sigma(r) &=& 4200 \frac{\mbox{g}}{\mbox{cm}^2} \left(\frac{r}{\mbox{1 AU}}\right)^{-3/2}.
\end{eqnarray}
  This provides  a natural scale for $D$  as a function of the local  disk radius and the central  stellar mass
through 
\begin{eqnarray}
D_0 &=& 2 C F_0^2  \tau_{c,0} \ \ 
= 25 \; C \; \frac{ \mbox{cm}^2}{\mbox{s}^{3}} \;\left( \frac{r}{1\mbox{ AU}}\right)^{-\frac32} \left(\frac{M_*}{1 \mbox{ M}_\odot}\right)^{-\frac 12},
\end{eqnarray}
where $C$ is a dimensionless constant.

There are several very uncertain factors which contribute to  determining an  appropriate  value of the dimensionless constant~$C$: 
The density fluctuations found in   MRI simulations  are typically 
${\delta \rho}/{\rho} = {\delta \Sigma}/{\Sigma} \approx 0.1$ \citep[eg.][]{Nelson2005}.  
The presence of a dead zone in the mid plane regions  of the disk,
 where the MRI is not active, has been found to cause  reductions in  the magnitude of
  $F_0$ by one order of magnitude or more,
as compared to  active cases \citep{oishi2007}.
Massive planets open a gap in the disk.
\cite{oishi2007} found that most of the contribution to the stochastic force
comes from density fluctuations within a distance of one scaleheight from the planet. 
When a gap forms, this region is cleared of material leading to the expectation of a  substantial decrease 
 in the  the magnitude  of turbulent density fluctuations. Consequently  $F_0$  should be  reduced on account of a lower ambient surface density. A factor of $\frac 1 {10}$ seems reasonable although it might be even smaller \citep{DiscComp2006}. 
 The correlation time $\tau_c$ is actually found to be  
approximately~$0.5\Omega^{-1}$ \citep{NelsonPapaloizou04,oishi2007}.

If it is appropriate to include  reduction factors to account for all of the above effects, one 
finds $C= 5\cdot 10^{-7}$ and  we expect a natural scale for the diffusion coefficient  to be
specified through 
\begin{equation}
 D_0 \rightarrow 10^{-5}\left( \frac{r}{1\mbox{ AU}}\right)^{-3/2}
  \left(\frac{M_*}{1 \mbox{ M}_\odot}\right)^{-1/2} \frac{ \mbox{cm}^2}{\mbox{s}^{3}}.\label{eq:scalingd} 
\end{equation}
The same value of D may be equivalently scaled to the orbital parameters
of the planets without reference to the disk by writing
\begin{eqnarray}
 D_0 &=& 3.5 \cdot 10^{6}\left( \frac{r}{1\mbox{ AU}}\right)^{-5/2}\left( \frac{M_*}{1M_{\odot}}\right)^{3/2} \nonumber \\
 && \quad \cdot \left(\frac{ r^4 \left< F_i^2 \right>\Omega \tau_c}{(GM_*)^2}\right) \frac{ \mbox{cm}^2}{\mbox{s}^{3}}. \label{DCOEFF}
\end{eqnarray}
Thus a value  $D_0=10^{-5}$ in cgs units corresponds to a ratio of the  root mean square stochastic 
force component to that due to the central star  
 of about $\sim  10^{-6}$ for a central solar mass at $1$~AU.
 It is a simple matter to scale to other locations.

Of course we emphasize  that the value of this quantity is very uncertain, a situation
that is exacerbated by its proportionality to the square of the magnitude of the stochastic
force per  unit mass. For this reason we perform simulations for a  range of $D$
covering many orders of magnitude.

\subsection{Numerical Implementation}
 The procedure we implemented, uses a discrete first order Markov process to generate a correlated noise that is continuous and added as an additional force. The Markov process is a statistical process which is defined by two parameters, the root mean square of the amplitude and the correlation time $\tau_c$ \citep{KASDIN}. 
  It has a zero mean value and has no memory. This  has the advantage that  previous values 
  do not need to be stored. The autocorrelation function decays exponentially and thus mimics the  autocorrelation function  measured in MHD simulations by \cite{oishi2007}.

\subsection{Stochastic Forces Acting on a Single Planet}
We first investigate the long term effect of stochastic  forces on a single isolated planet.
 The initial orbital parameters were taken to be the the observed parameters of GJ876~b (see table \ref{table:initorb}) and the central star had a mass of  $0.38~\mbox{M}_{\odot}$.
In this simulation, we use the reduced \changed{diffusion coefficients}
\begin{eqnarray}
 D &=& 8.2 \cdot 10^{-5} \frac{ \mbox{cm}^2}{\mbox{s}^{3}} \quad \text{and}\\
 D &=& 8.2 \cdot 10^{-3} \frac{ \mbox{cm}^2}{\mbox{s}^{3}} 
\end{eqnarray}
which can be represented by a correlation time of half the orbital period  
 and a specific force with root mean square values $\sqrt{\left<F^2\right>}~=~4.05\cdot10^{-6}~\mbox{cm}/\mbox{s}^{2}$ and $\sqrt{\left<F^2\right>}~=~4.05\cdot10^{-5}~\mbox{cm}/\mbox{s}^{2}$, respectively
 (see also equation (\ref{DCOEFF}) above).
  
\begin{figure}
\centering
\includegraphics[width=\columnwidth]{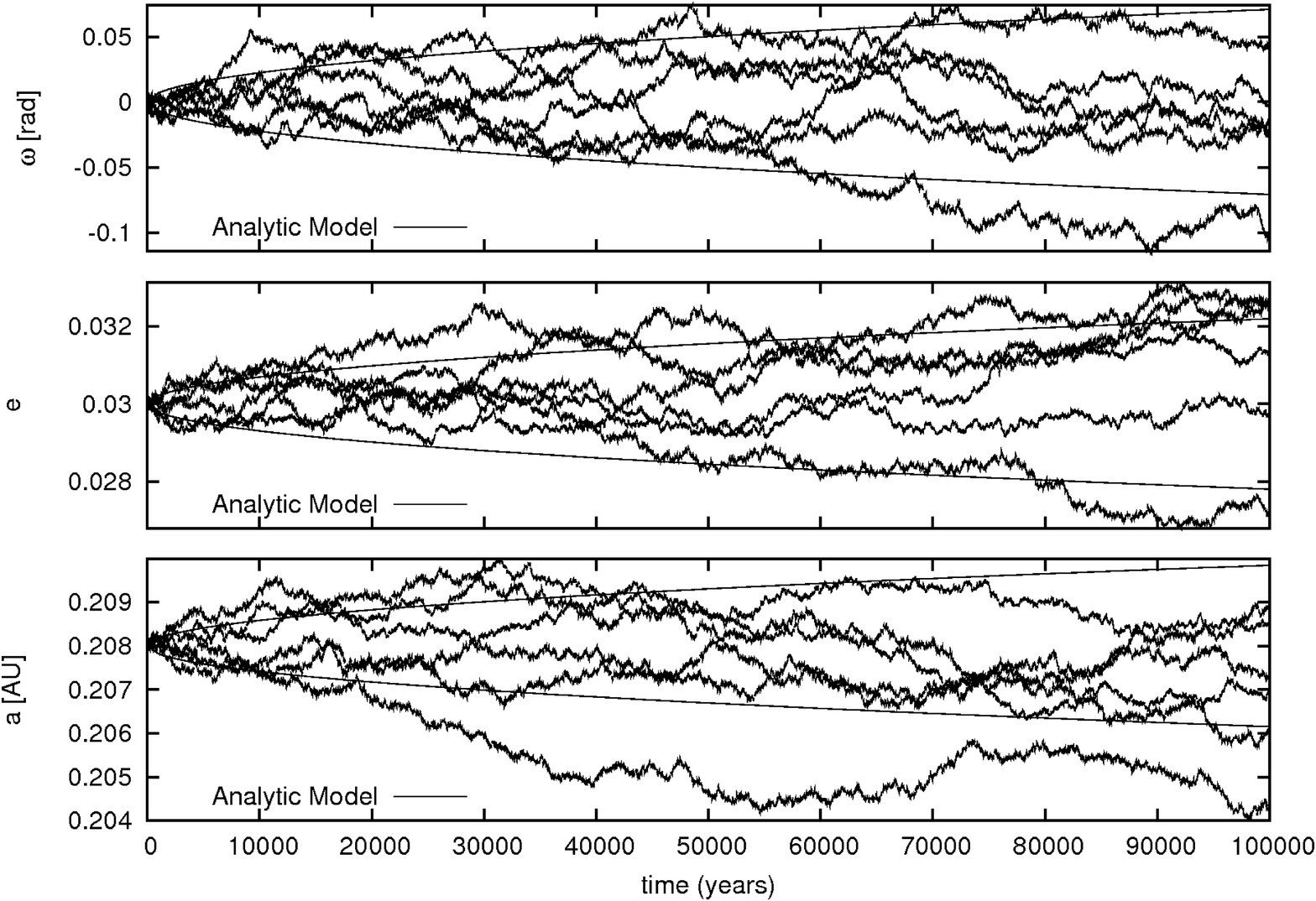}
\includegraphics[width=\columnwidth]{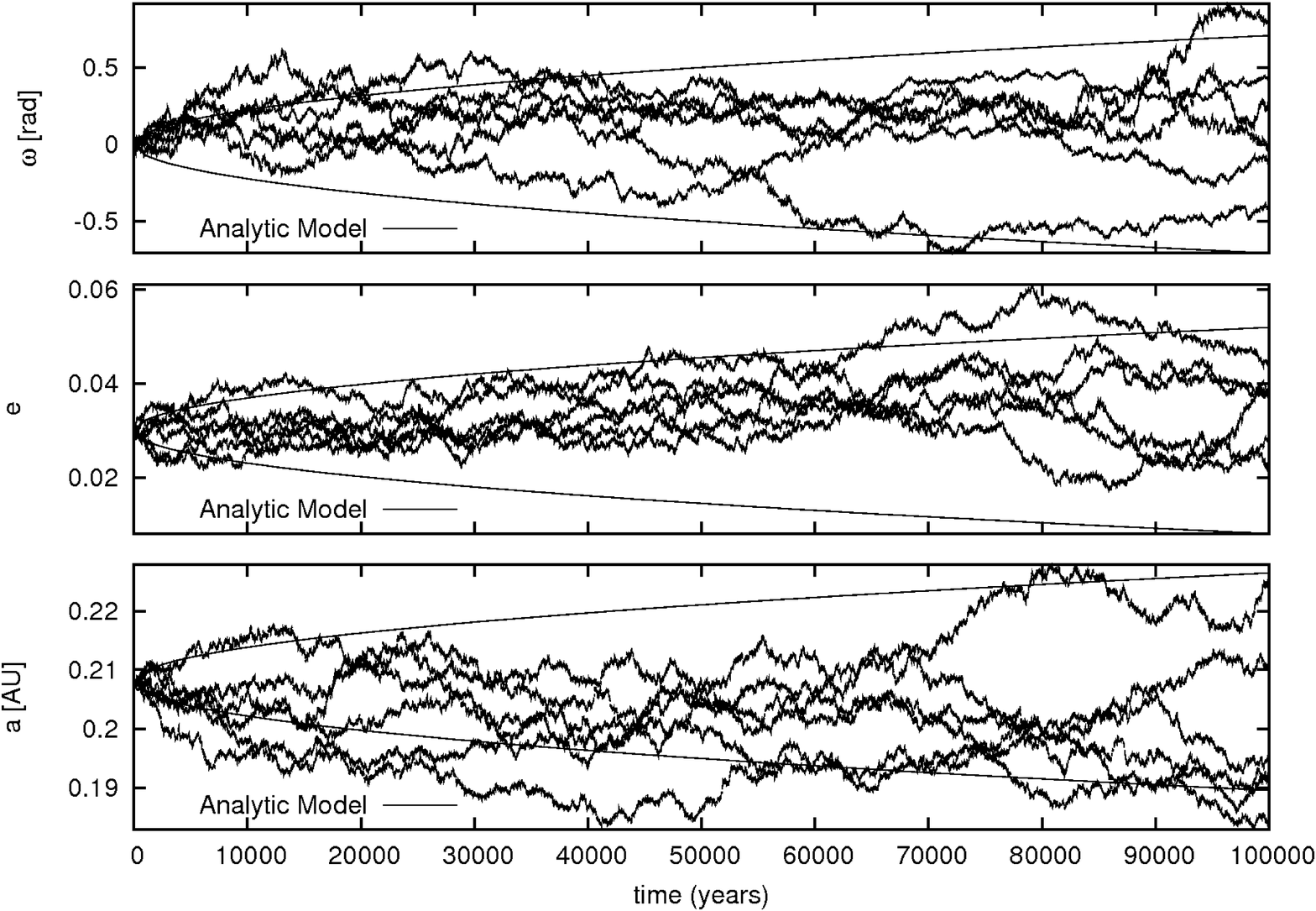}
\caption{Time evolution of the longitude of periastron $\varpi$, the eccentricity $e$ and the semi major axis $a$   for a single planet. The initial orbital parameters of the planet  were taken to be 
those of GJ876~b  (see table \ref{table:initorb}). 
Six different realisations starting from
 the same initial conditions are shown in each panel. 
The diffusion coefficients are  $D = 8.2 \cdot 10^{-5} { \mbox{cm}^2}/{\mbox{s}^{3}}$  for  the upper three panels
  and $D = 8.2 \cdot 10^{-3} { \mbox{cm}^2}/{\mbox{s}^{3}}$  for 
the lower  three  respectively. 
The solid lines  correspond to  the analytic predictions for the amount of spreading
  (see text). }
 \label{fig:singleplanet}
\end{figure}

The resulting random walks undergone by  $e$, $a$ and $\varpi$ are plotted in figure \ref{fig:singleplanet} for six different realisations for each of the two diffusion parameters.  
The spreading rates can be estimated from equations (\ref{eq:growtha}) - (\ref{eq:growthw}). These analytic predictions are plotted as solid lines. 
Clearly the numerical model is in broad agreement with the random walk description with a spreading that scales with $D$ as expected. 
We have performed simulations for a variety of $D$ and get  results that are fully consistent
with this  in all cases.  

\subsection{Illustration of the Modes of Libration in a Two Planet System}
We now  go on to consider two planet systems.
As an illustrative example we consider the GJ876 system 
(see table \ref{table:initorb}). Note that we can easily scale all physical quantities and extend the discussion to other systems (see section \ref{sec:discussion} for a discussion).
We begin by considering the evolution of the system
without stochastic forcing in order to  characterize the 
modes of libration of the resonant angles and other orbital parameters.
In particular we identify the fast and slow modes discussed in
section \ref{FSmodes}.

\begin{figure}
\centering
\includegraphics[width=\columnwidth]{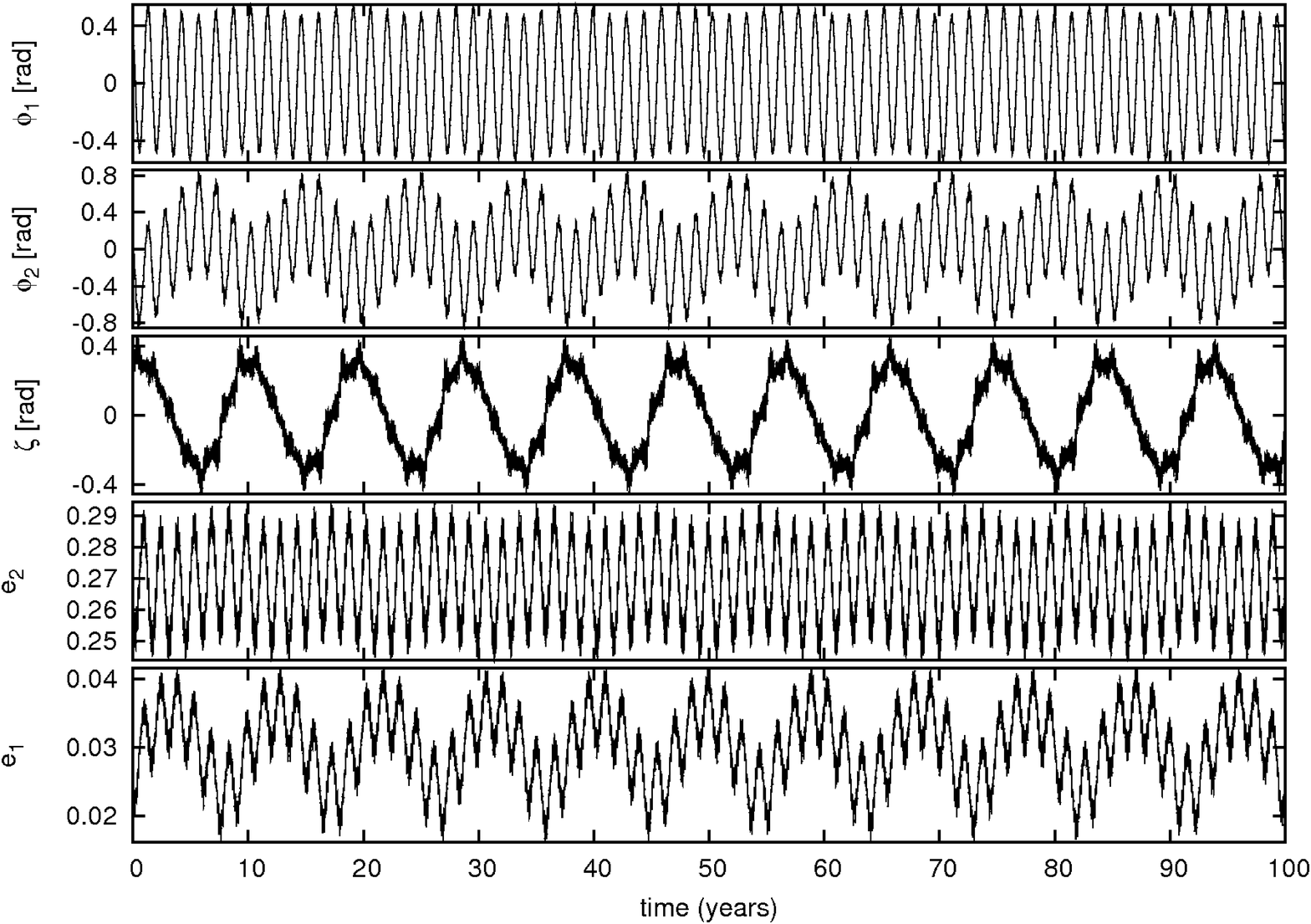}
\caption{Time evolution of the resonant angles and the eccentricities in the system GJ876 without  turbulent forcing. The dominance of the  fast
mode with period $\sim 1.4$~years in the oscillations
of $\phi_1$,  and the dominance of the   slow mode 
with period $\sim 10$~years in $\zeta$  can be clearly seen. }
 \label{fig:twoplanets}
\end{figure}

The time evolution of the resonant angles and the eccentricities 
is plotted in figure \ref{fig:twoplanets}.
Clearly visible are the  slow and fast oscillation  modes. 
The fast mode, which  is seen to have a  period of about 1.4 years,
  dominates    the librations of  $e_2$, and  $\phi_1$ 
while  also  being present  in those of $\phi_2.$ 
On the other hand the slow mode,  which  is seen to have a 
 period of about 10 years,
 dominates  the librations of 
$\zeta$ while  also being present
  in those of $e_1$ and $\phi_2.$

We emphasize  the fact that the eccentricities of the 
two planets participate in the librations and so  are not constant. 
  In particular, the eccentricity of GJ876~b oscillates around a mean value of $0.03$ with an amplitude $\Delta e \approx 0.01 - 0.02$. 
This behaviour
involving the attainment of smaller values of the eccentricity
has important consequences for   stochastic evolution  as discussed 
 above (see section \ref{GRLIB})  and see  also  below. 

We remark that similar behaviour occurs for all the systems 
we have studied, these have a wide range of eccentricities
and planet masses. Indeed we note that the period scalings 
of the fast and slow mode periods  with the planet masses 
are provided by equations (\ref{fastp}) and (\ref{slowp}) respectively.
These indicate, as confirmed by our simulations,
 that if both masses are reduced by a factor $\Lambda$
then the period of the fast mode scales as $\sqrt{\Lambda}$
and the period of the slow mode scales as $\Lambda.$

\section{Two Planets with Stochastic Forcing}
We now consider systems of two planets with stochastic forcing.
For simplicity we begin by applying forcing only to the outer planet.
We have found that adding the same form of forcing
to the inner planet tends to speed up the
evolution by approximately a factor of 
two without changing qualitative details.
For illustrative purposes we again start
with the GJ876 system  and consider two  diffusion coefficients 
$D = 8.2~\cdot~10^{-5}  \text{cm}^2/\text{s}^{3}$ and $D = 8.2~\cdot~10^{-4}  \text{cm}^2/\text{s}^{3}$. 
We remark that all of our simulations were run with constant 
values of $D=2\langle F_i\rangle ^2\tau_c.$
This was done maintaining  the parameters $\left< F_i\right>$ and $\tau_c$ to be  constant, with $\tau_c$ being determined 
for the initial location of the outer planet.
For the cases considered here, there is little orbital migration
so this is not a significant feature.
Note also  that 
different realizations of the  system
are likely to become unstable and scatter for higher values of $D$ (see below).

\begin{figure}
\centering
\includegraphics[width=\columnwidth]{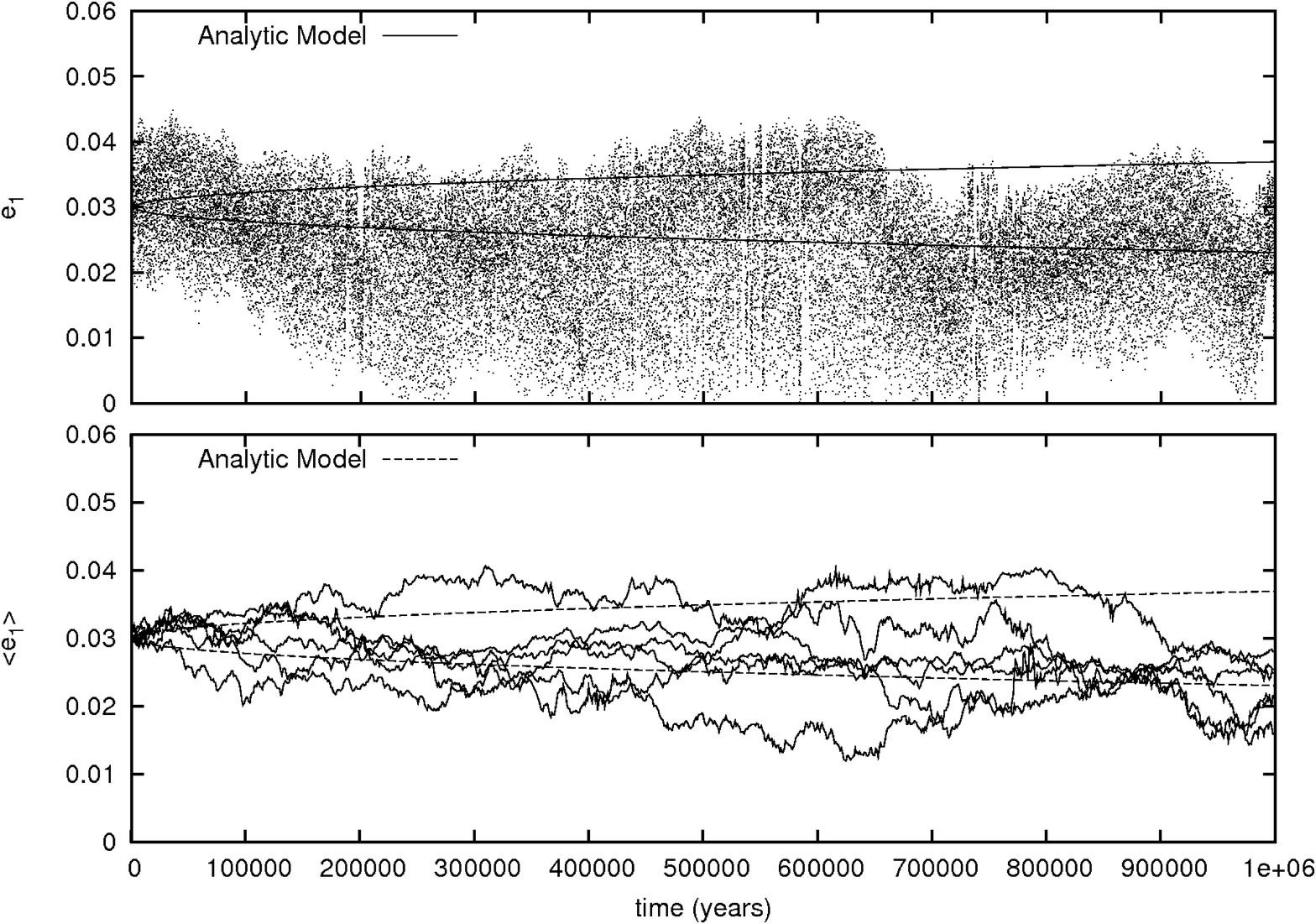}
\includegraphics[width=\columnwidth]{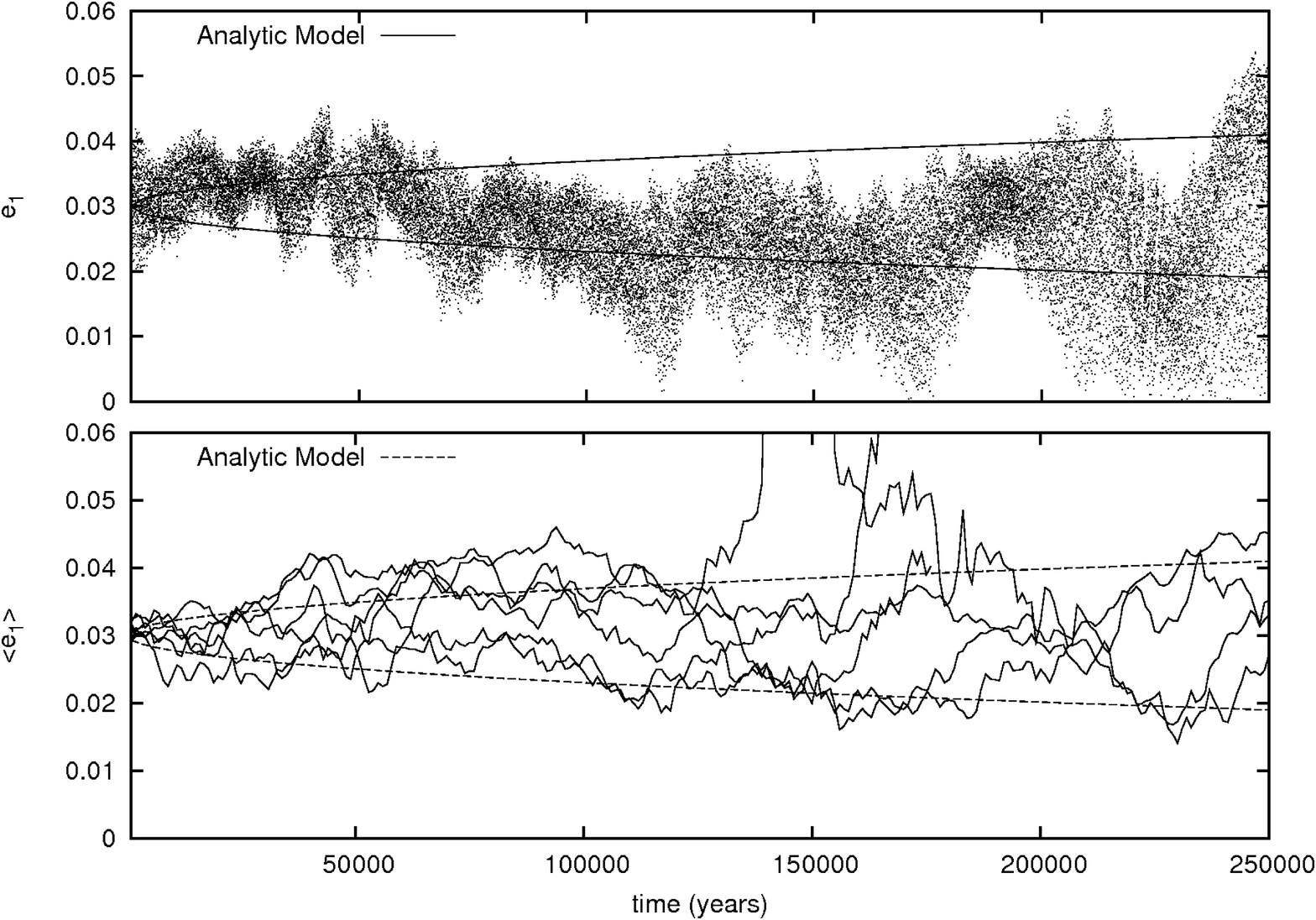}
\caption{Time evolution of the eccentricities 
in the GJ876 system with  turbulent forcing
included. The diffusion coefficient is $D = 8.2~\cdot~10^{-5} \text{cm}^2/\text{s}^{3}$ for the two upper   plots and $D = 8.2~\cdot~10^{-4} \text{cm}^2/\text{s}^{3}$ for the two lower  plots. The first (uppermost) and the third plots each show a single run. The second and fourth (lowermost) plots show the time averaged eccentricity for  four runs. The averaging interval is 1000~years. }
 \label{fig:twoplanets_turb}
\end{figure}

The time evolution of the eccentricities is plotted 
in figure \ref{fig:twoplanets_turb}. 
We see  fast  oscillations superimposed on a random walk. 
The amplitude of the oscilations, as well as the mean value,
 changes with time.  

Our simple analytic model, assumes slowly changing background
eccentricities and semi-major axes and   accordingly does not incorporate the 
oscillations of the eccentricity  due to the resonant 
interaction of the planets. 
In order to make a comparison, 
we perform  a  time   average  over many  periods  
to get smoothed quantities whose behaviour
 we can compare with that expected from 
 equations (\ref{eq:growtha}) - (\ref{eq:growthw}). 
 When this procedure is followed,
 the evolution is in reasonable accord  
with that expected from the analytic model provided 
that allowance is made for the importance of small values of $e_1$ in
 determining the growth of the libration amplitude
of the  angle between the apsidal lines of the two planets
 (see equation (\ref{eq:growthw})).
The presence of this feature results in the behaviour of the libration amplitude being more complex than that implied by a process
governed by  a  simple diffusive random walk. 

\subsection{ Disruption of a Resonance in Stages}
Systems with mean motion commensurabilities can be in many different configurations. Here we describe
the important evolutionary stages 
as they appear in a stochastically forced system 
that starts  from a situation in which all the resonant
angles show small amplitude libration. 
For example
observations of GJ876 suggest that the system is currently  in such a state
 with the ratio of the orbital periods $P_1/P_2$  oscillating 
 about a mean value of $2.$

\subsubsection {Attainment of Circulation of the Angle
Between the Apsidal Lines}
When the initial eccentricity  of the outer planet, $e_1,$
is small the excitation of the libration amplitudes of the resonant
angles readily brings about a situation where $e_1$ attains very small values.
This  causes  the periastron difference  $\zeta$
 to undergo large oscillations 
and eventually circulation (see equation (\ref{eq:growthw})).
Should stochastic forces cause
   $e_1$  to reach zero, \changed{ on account of the coordinate singularity} $\varpi_1$  becomes undefined.
  Subsequently very small perturbations  are able to produce  a
 small eccentricity
 with
 $\zeta$ undergoing  large amplitude 
librations or circulation (see below).
\changed{But note that as also mentioned in section \ref{AdF}, this occurs without a large physical
perturbation to the system.
 Thus} the occurence of
  this phenomenon  does not imply the system ceases to be in a commensurability.

In this context we note that the eccentricity of GJ876~b 
initially is such that $e_1 \sim 0.03$ 
with values $e_1\sim 0.01$  often being attained during libration cycles.
Thus only a small change may cause the above situation to occur.
In all cases we have considered, 
 we find that  $\zeta$ enters circulation 
prior to the fast angle  $\phi_1,$ which may remain librating until that
too is driven into circulation.

\subsubsection{Attainment of Circulation of the Fast Angle}
Both before and
after $\zeta$ enters circulation, stochastic forcing 
 acts to increase the libration amplitude of the fast mode.
 This mode dominates both 
 the librations of the resonant angle $\phi_1$ and the semi major axes.
  Eventually $\phi_1$ starts circulation. 
Shortly afterwards commensurability is lost and $P_1/P_2$ starts 
to undergo a random walk with a centre that drifts away from 2.
   Note that it is possible for 
some realizations to re-enter commensurability.  For systems
   with  the masses of the observed GJ876 system, 
the most likely outcome is a scattering event
   that causes complete disruption of the system.

\subsubsection{A Numerical Illustration}
\begin{figure}
\centering
\includegraphics[width=\columnwidth]{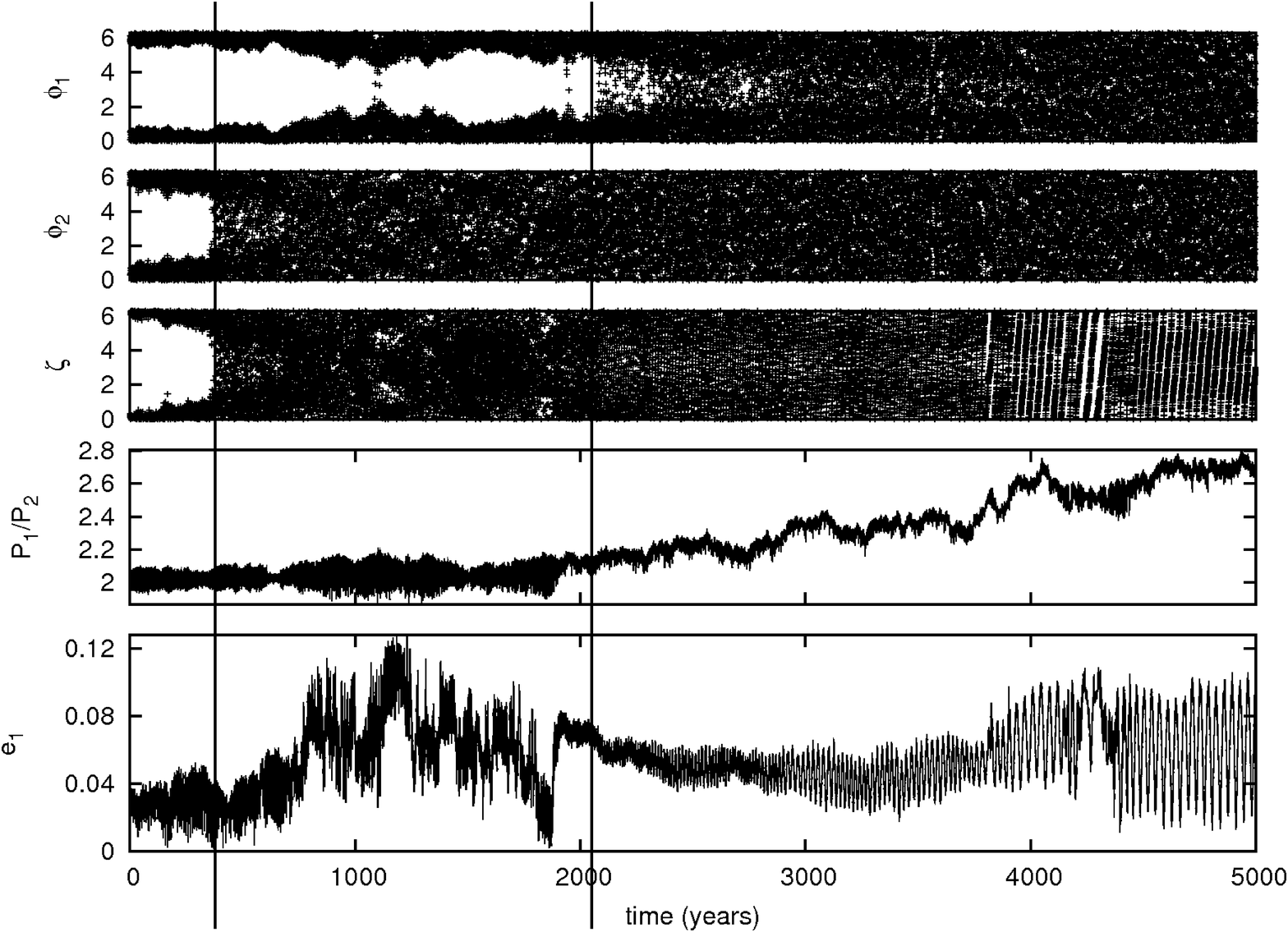}
\includegraphics[width=\columnwidth]{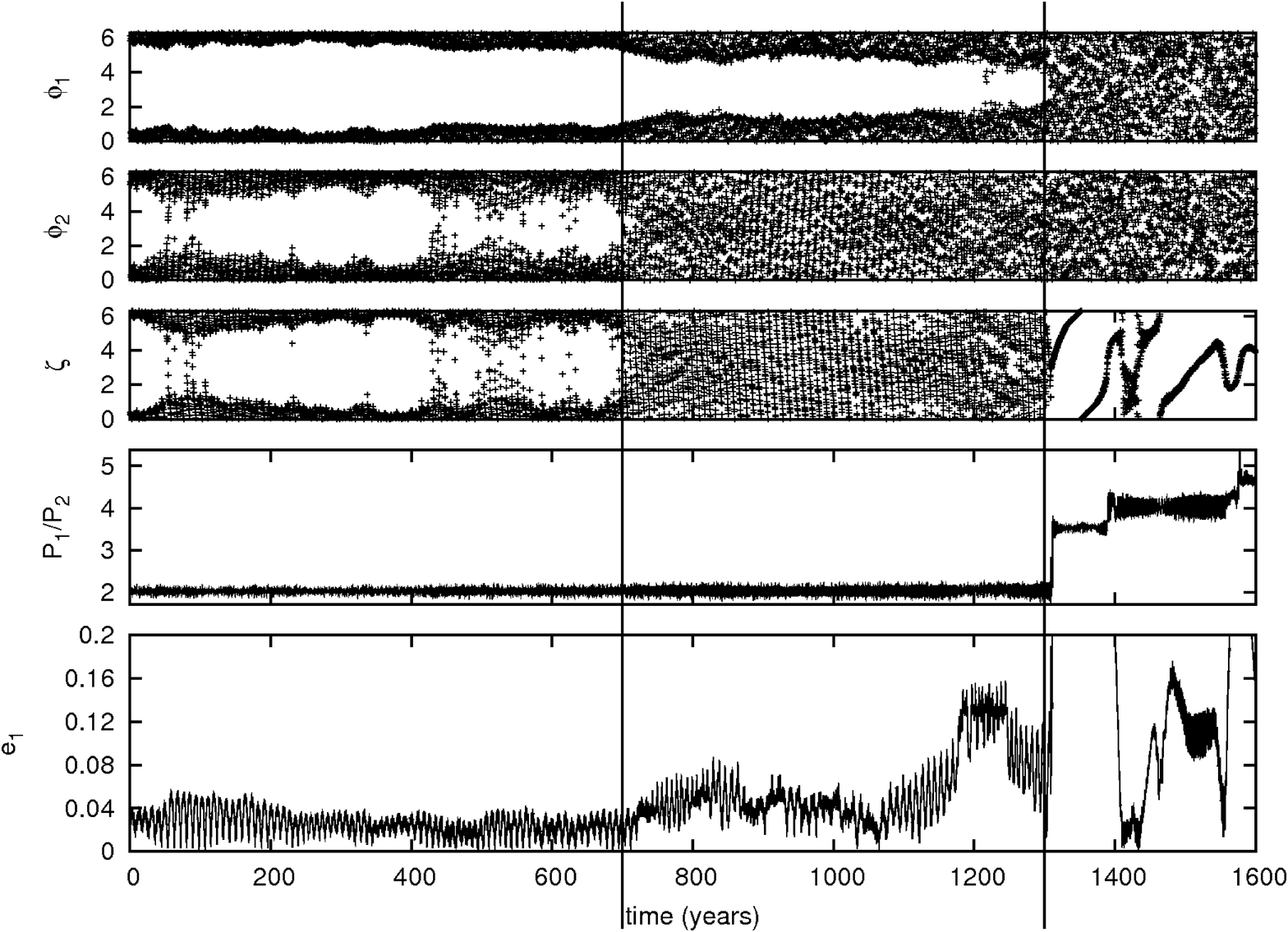}
\caption{Time evolution of the resonant angles, 
the period ratio $P_2/P_1$ and the eccentricity, $e_1,$
 in the  GJ876 system with  stochastic forcing corresponding to 
  $D = 0.42 \text{cm}^2/\text{s}^{3}$. 
The vertical lines indicate when the angles enter circulation
for a prolonged period.
 The realisation illustrated in the lower panel
  scatters shortly after $\phi_1$ goes
 into circularization. }
 \label{fig:twoplanets_turb_break}
\end{figure}

In order to illustrate the 
 evolutionary sequence described above we plot results
  for two realisations of the evolution of the GJ876 system
in figure \ref{fig:twoplanets_turb_break}.
  For these runs we adopted  the diffusion coefficient
$D = 0.42 \text{cm}^2/\text{s}^{3}.$ 
In this context we note that reducing $D$ increases
the evolutionary time which has been found, both analyticaly and numerically 
 to be~$\propto~1/D$ (see below).

The times at which the transition 
from libration to circulation occurs for both the slow and fast angles are
indicated by the vertical lines in figure \ref{fig:twoplanets_turb_break}.
Several of the features discussed above and in section \ref{GRLIB}
can be seen in figure \ref{fig:twoplanets_turb_break}.
In particular the tendency for the occurence of very small
values of $e_1$ to be associated with transitions to circulation
of  $\zeta$ can be seen for the realisation plotted in the lower
panels at around $t=80$~years and $t= 500$~years.
Such episodes always seem to occur when the libration amplitude
of the fast angle $\phi_1$ is relatively boosted, 
indicating that this plays a role
in boosting the slow angle. If the period of time for which $e_1$ attains
small values is small and $\phi_1$ recovers small amplitude librations,
the slow angle returns to circulation.
Thus the attainment of long period circulation for the slow angle
is related to the diffusion time for $\phi_1.$
We also see from  figure \ref{fig:twoplanets_turb_break}
that the angle $\phi_2,$ which  has a large contribution from  the slow mode,
 behaves in the same way as $\zeta $
as far as libration/circularization is concerned.
We have verified by considering the results from the simulations of
GJ876 LM HE,  which started with a larger  value of $e_1,$ that, as expected,
the attainment of circulation of the slow angle takes relatively longer
in this case, the time approaching more closely the time when $\phi_1$
attains circulation. 
Also as expected, the time when $\phi_1$ attains circulation is not affected by the change in $e_1$.

\subsection{Dependence on the Diffusion Coefficient}
We now consider  the stability of the systems 
listed in  table \ref{table:initorb} as a function of $D.$
These systems have a variety of masses and orbital eccentricities.
In particular, in view of the complex interaction between
the resonant angles discussed above we wish to  investigate
whether the mean amplitude growth at a given time is indeed  proportional to
$D.$ 
As also  mentioned above, the value of the diffusion parameter $D,$
that should be adopted,  is very uncertain. 
We have therefore   considered values of $D$ ranging 
over five orders of magnitude. 
However, the correlation time $\tau_c$ is always taken 
to be given by  $\tau_c = 0.5\Omega^{-1}$ while  the RMS value of the force
is changed. 
In order to determine 
the "lifetime" of a resonant angle, we monitor whether 
it is librating  or circulating.  Numerically, libration is defined
to cease at  the first time the angle is seen to 
reach absolute values largen than $2$. 
We note that the angle can in general regain small values  afterwards. However, this is  a transient effect and changes the lifetime by no more than a factor of $\sim 2$ in all our simulations.

In this context we consider the fast angle $\phi_1$ and the slow angle $\zeta,$
though, as we saw above,  the latter can be replaced by $\phi_2$
which exhibits the same behaviour. 
As $\phi_1$ is the last to start circulating
the resonance is defined to be broken at that point.

\begin{figure}
\centering
\includegraphics[width=\columnwidth]{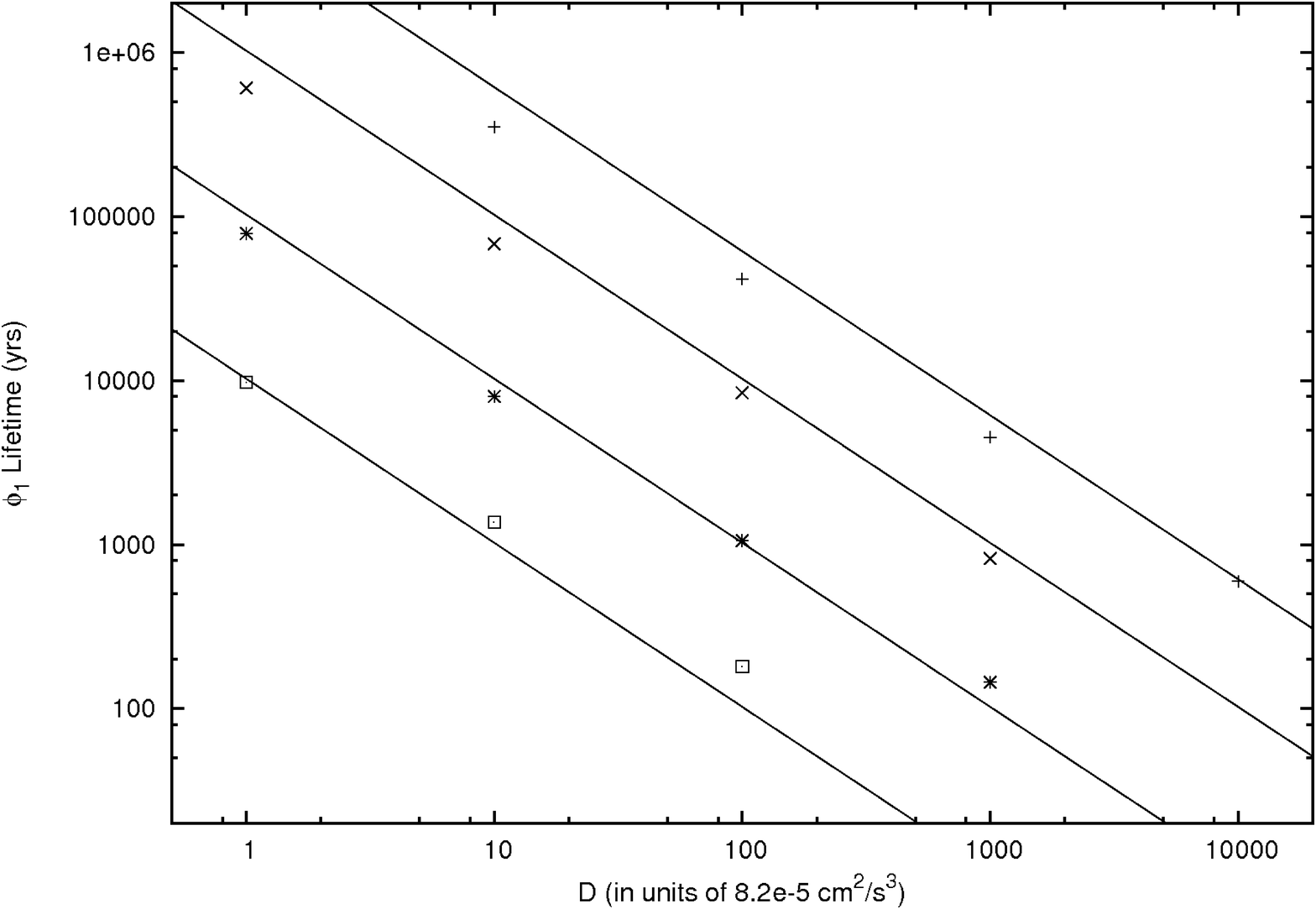}
\includegraphics[width=\columnwidth]{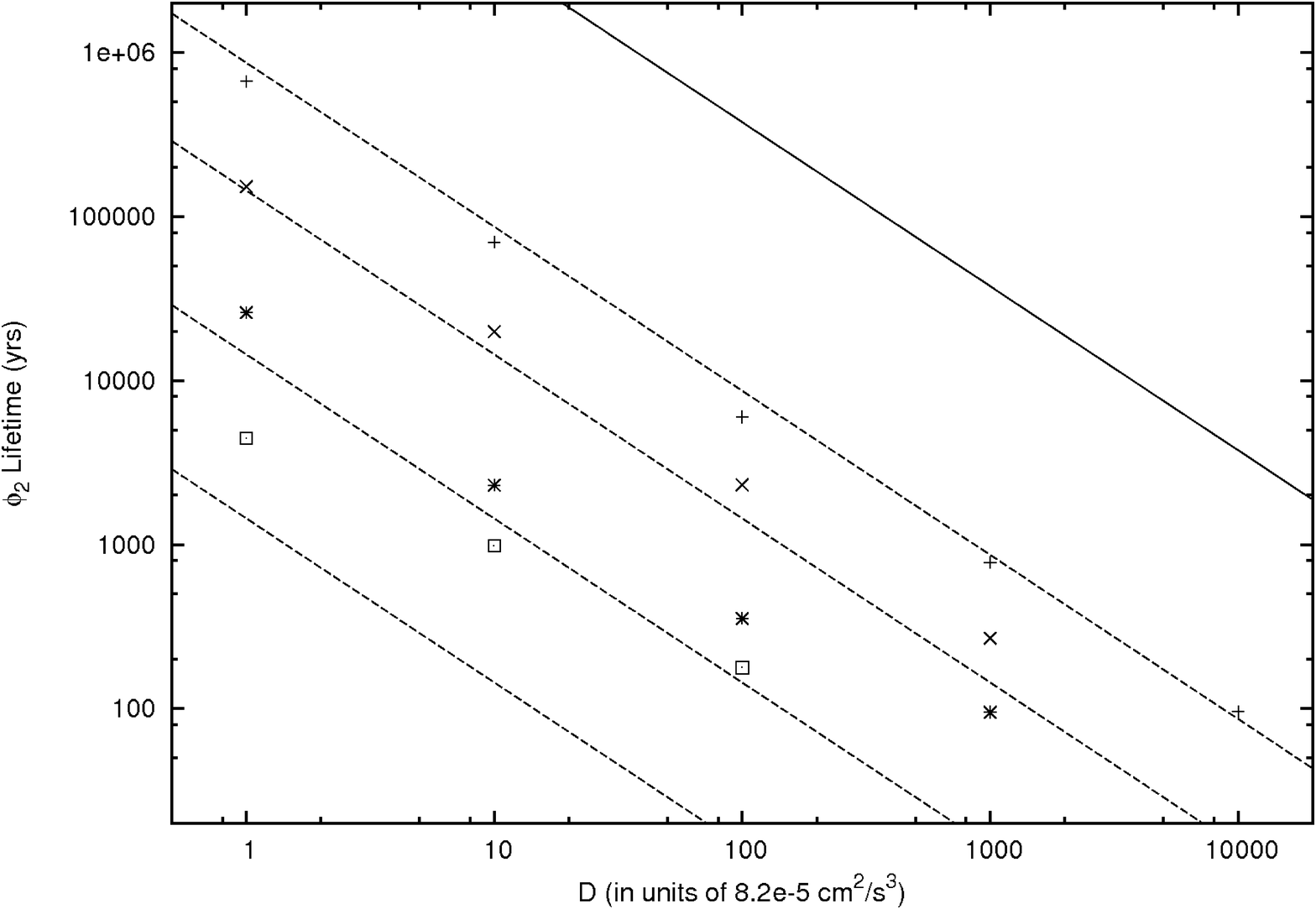}
\caption{Average time until circulation of the resonant angles $\phi_1$ (top) and $\phi_2$ (bottom) in the GJ876
 (indicated with $+$) , GJ876LM (indicated with $\times$), GJ876SE (indicated with $\star$)  and GJ876E (indicated with $\Box$)
systems  
as a function of the stochastic forcing diffusion coefficient $D.$
In the case of the upper panel, the analytic curves explained in the text,  are from top to bottom for the 
GJ876, GJ876LM, GJ876SE and GJ876E 
systems respectively. In the lower panel the analytic curves, as explained in the text,  are from top to bottom for the GJ876 
 (solid curve and top dashed  curve), GJ876LM, GJ876SE and GJ876E
systems respectively.   }
 \label{fig:lifetime1}
\end{figure}

Equations (\ref{eq:growthf}) and (\ref{eq:growths}) 
estimate the spreading of the resonant angles
as a function of time.  We determine the times to attain circulation
as the times to attain $(\Delta \phi_i)^2 = 4, i = 1,2 $
\changed{assuming the initial values for the orbital elements.} 
We plot both the numerical and analytical 
results in figure \ref{fig:lifetime1}. 
To remove statistical fluctuations 
and obtain a mean spreading time, the numerical values  for a particular value of $D$ were obtained by averaging over $60$ realizations.

From figure \ref{fig:lifetime1} it is apparent that the evolutionary times scale is $\propto 1/D$
for $D$ varying by many orders of magnitude. The analytic estimates for the libration 
survival times of $\phi_1,$
dominated by the fast mode,   obtained using equation (\ref{eq:growthf})
\changed{adopting the initial orbital elements and }  with the
fast libration frequency
determined from the simulations, are plotted in the upper panel
of figure  \ref{fig:lifetime1}. These 
are  in good agreement with the numerical results.
However, the estimate for $\phi_2$ based on  equation (\ref{eq:growths}) using the initial value
of $e_1$ overestimates the lifetime by a factor of at least 
 $\sim 40$ (see solid line in the lower plot of figure \ref{fig:lifetime1}).
Futhermore (\ref{eq:growths}) has no dependence on planet mass
which is in clear conflict with the numerical results.
  This, as discussed above is due
to the temporary attainment of small eccentricities.  This causes the disruption of the libration
of $\zeta$ earlier than would be predicted assuming $e_1$ is constant. In fact this disruption
occurs at times  that can be up to a factor of $10$ times shorter than those required 
to disrupt $\phi_1$. We  estimate the lifetime 
of librations of  $\phi_2$ by calculating the time $\phi_1$ needs to reach values close to $1$ (see dashed lines in the lower plot of figure \ref{fig:lifetime1}).
 As  explained  above in section \ref{GRLIB}, large amplitude variations of $\phi_1$
are expected to couple to and excite  
the slow mode.  Variations   induced in the eccentricity $e_1$ allow this to reach
 zero and we lose libration of $\phi_2.$ 
  These simple estimates are in good agreement with the numerical results and
accordingly support the idea that $\phi_1$ and $\phi_2$ are indeed coupled in the non linear regime.
 For low mass planets this simple  analytic prediction underestimates the lifetime of $\phi_2$
  by a factor of $\sim 2$, suggesting that the coupling between the two modes depends slightly on the planet masses.

To   confirm our understanding of these  processes, we have repeated the calculations 
with systems that are in the same state as the system discussed above but have higher 
eccentricities (see GJ876~LM~HE and GJ876~SE~HE in table \ref{table:initorb}). The lifetime of  the  librating
state of  $\phi_1$ is unchanged, as this does not depend significantly  on the eccentricity. 
However, the lifetime of the librating state  of  $\phi_2$ is a factor of $2-3$ longer. 
This is due to the fact that a larger excitation of $\phi_2$ is needed to make $e_1$ reach small values
in these simulations. 

The random walk description breaks down completely  when the anticipated disruption time
becomes shorter than the libration period. This is expected to happen
for small enough masses as can be verified from figure \ref{fig:lifetime1}.
It is because the disruption time decreases linearly with the planet masses
while the libration period increases as the square root of the planet masses.
Then we cannot average  over many libration periods.
This situation is apparent in figure   \ref{fig:lifetime1}
    for very short disruption times of the order 100-1000~years where survival times
cease to vary linearly with $1/D$.

\section{Formation of HD128311}

We here discuss the application of the ideas discussed above to understanding
the orbital configuration of 
the planetary system HD128311.    This is (with 99\% confidence)  in a 2:1 mean motion resonance with the angle $\phi_1$  librating  and the angle $\zeta$ circulating
so that there is no apsidal corotation \citep{Vogt2005}. 
However,  the orbital parameters are not well constrained. The orgininal Keplerian fit by \cite{Vogt2005} is such that the planets undergo a close encounter after only 2000 years. The values in table \ref{table:initorb} have been obtained from a fit to the
observational data that includes interactions between the planets. 
The values quoted have large error bars. For example the eccentricities $e_1$ and $e_2$  have an uncertainty of 33\% and 21\%, respectively. Although the best fit doesn't manifest apsidal corotation, the system could undergo large amplitude librations  and still be stable.

According to \cite{LeePeale2002} the planets should have apsidal corotation if the 
commensurability was formed  by the two planets undergoing sufficently slow 
convergent inward  migration, and if they then both migrated inwards significantly  
while maintaining the commensurability \citep[see also][]{SnellgrovePapaloizouNelson01}.
Accordingly  \cite{SandorKley06} suggested a possible formation scenario with inward
migration as described above, but with an additional perturbing event, such as  a close encounter with an additional planet occuring after the halting of the inward migration. This perturbation is needed to alter the behaviour of  $\zeta,$ 
so that it undergoes circulation rather than libration
and thus producing orbital parameters similar to the observed ones. 

We showed above that stochastic forcing possibly resulting
from  turbulence driven by the MRI readily produces systems with commensurabilities
 without apsidal corotation if the eccentricities are not too large. 
 This suggests that a scenario that froms the commensurability
 through disk induced inward convergent migration might readily produce commensurable systems
 without apsidal corotation if stochastic forcing is included.
Such scenarios are investigated in  this section.

\begin{figure}
\centering
\includegraphics[width=\columnwidth]{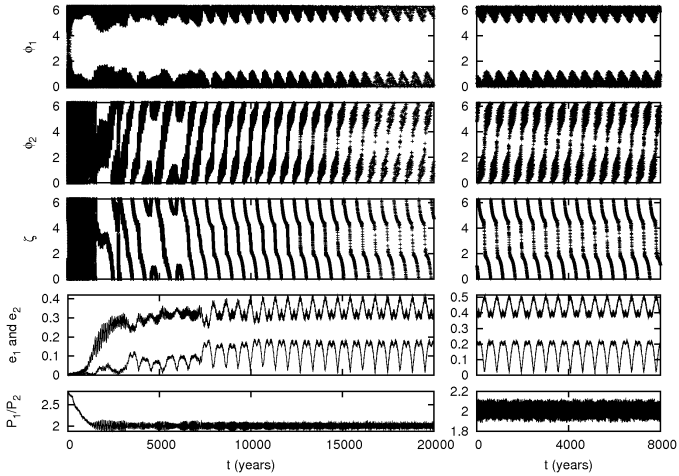}
\caption{
The plots on the left hand side show a possible formation scenario of HD128311. 
We plot the observed system on the right hand side as a comparison (see table~\ref{table:initorb} and text). 
The plots show the resonant angles $\phi_1, \phi_2, \zeta$, the eccentricities $e_1, e_2$ and period ratio $P_1/P_2$ for formation scenarios including turbulence and migration. 
Resonace capturing occurs after 2000~years.
 }
 \label{fig:hd128311}
\end{figure}

\begin{figure}
\centering
\includegraphics[width=\columnwidth]{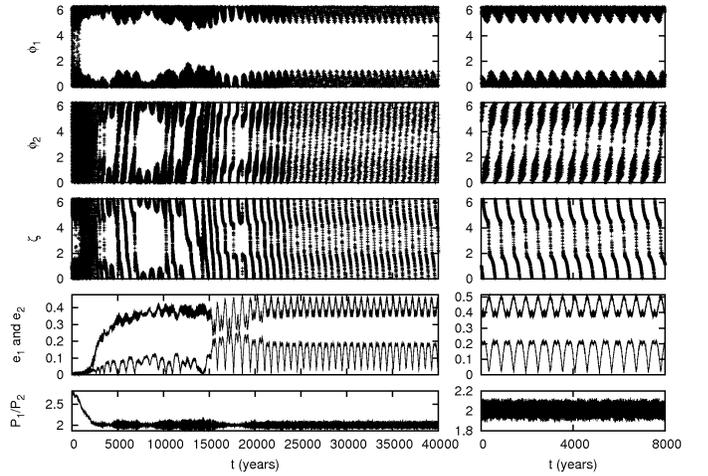}
\caption{
The plots on the left hand side show another possible formation scenario of HD128311. 
Again, we plot the observed system on the right hand side as a comparison (see table~\ref{table:initorb} and text. 
The migration timescales in this run are  $\tau_{a,1}=16000$, $\tau_{a,2}=-40000$ and $K=5.5$. Note that the eccentricities are larger compared to figure \ref{fig:hd128311} because $K$ is smaller. 
 \label{fig:hd128311_2}
}
\end{figure}

\begin{figure}
\centering
\includegraphics[width=\columnwidth]{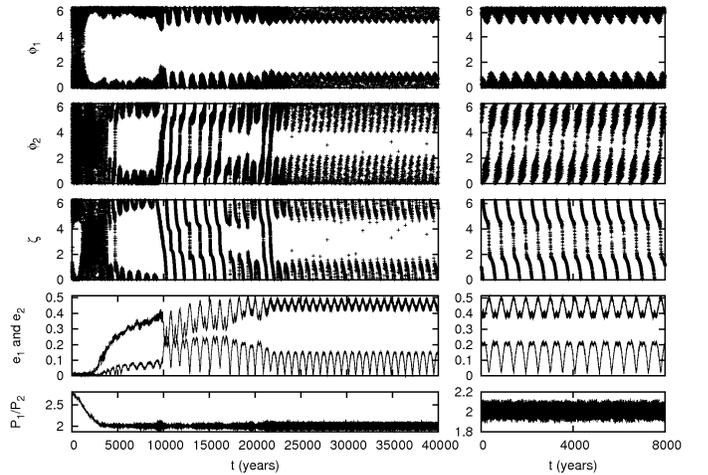}
\caption{
These plots show another possible formation scenario of HD128311. 
The damping parameters are the same as in figure \ref{fig:hd128311_2}.}
 \label{fig:hd128311_3}
\end{figure}

 We present simulations of the formation of HD128311 that include migration 
 and stochastic forcing
 due to turbulence but do not invoke special perturbation events
 involving additional planets.   We find that model systems
 with orbital parameters resembling the observed ones
 are readily produced that give better matches than  so far provided by \cite{SandorKley06}.
The planets in this system are of the  order of several Jupiter masses and the eccentricity of one planet can  get very small during one libration period. The observed orbital parameters are given in table \ref{table:initorb} and plotted on the right hand side of figures \ref{fig:hd128311}, \ref{fig:hd128311_2} and \ref{fig:hd128311_3}. The mass of the star is $0.8~{\rm M}_{\odot}$.

\subsection{Migration Forces}
We incorporate the  non-conservative forces  excerted by a protostellar disk that  lead to inward migration by  following  the approach of \cite{LeePeale01} and \cite{SnellgrovePapaloizouNelson01}.
In this the migration process is characterized by migration and eccentricity damping timescales, $\tau_a = {a}/{\dot a}$ and  $\tau_e = {e}/{\dot e}$. 
We also define the ratio of the above timescales to be $K=\left|\tau_a\right|/\tau_e$. This ratio determines the eccentricities in  the state of self similar inward migration
of the commensurable system that is attained after large times. 
We verify this result for two different values of $K$ (see below). 
The procedure is now widely used  \citep[e.g.][]{SandorKley06, Crida08}  and we have checked our code against the results of their work. 

In this work we allow the planets to form a commensurability through convergent inward migration
but stochastic forcing is included, the disk is then removed through having its surface 
\changed{density reduced to zero on a 
2000~year timescale. This  simultaneously reduces both the migration forces and the}
stochastic forcing to zero. This procedure is very similar to that adopted by
the above authors but we have included stochastic forcing and removed the disk on 
longer time scales.

The stochastic forces have to have the right balance  with respect to the migration rate.
We have found that inward migration imparts stability to the resonant
system.
 If the migration rate is too fast relative
 to the stochastic forcing the migration 
 keeps down  the libration amplitudes and we do not get circulation. 
 On the other hand large eccentricity damping favours broken apsidal corotation. This might look counterintuitive at first sight, but remember that the diffusion of $\zeta$ depends on $1/e^2_{1,2}$. 
 Due to the stochastic nature of the problem, it is hard to present a continuum of solutions so we restrict ourselves to the discussion of three representative examples.  
However, we comment that we are able to obtain similar end states for  a wide range
of migration parameters. Before presenting these examples we briefly indicate 
how the librations can be stabilized by the migration process.

As above, the parameters $\left< F_i\right>$ and $\tau_c$ are kept constant, so \changed{maintaining}
$D$ constant,  for the duration of the simulation, with $\tau_c$ being determined 
for the initial location of the outer planet. As discussed above, these values may scale with the radius of the planet and we thus expect them to change during migration. However, the semi major axis of the outer planet changes only by $\sim 30$~\% during the simulation
consequently we ignore this effect. 

\subsection {\changed{Adiabatic Invariance and the Stabilization of Librations through Damping}}
Both the fast angle $\phi_1$ and the slow angle  $\phi_2$
  obey   stochastically forced oscillator equations.
Although we do not have a simple expressions for the oscillation frequencies
we know that for fixed eccentricity and 
planet masses they both  scale as the orbital frequency, $n_1.$
When the planets migrate inwards together, this increases slowly with time.
Accordingly the theory of adiabatic invariance indicates that the
libration amplitude should scale as $n_1^{-1/2}$
\changed{  \citep{Peale76}.}
If there were no stochastic forcing or other effects,
 the libration amplitude would decrease on twice the migration
time scale $2\tau_a$ \changed{once steady eccentricities were formed.
This provides only a small effect given the typical amount
of migration $\sim 30\%$ in our simulations.}

\changed{A comparison of the behaviour of a model of the GJ876 system
undergoing convergent orbital migration resulting in the fomation of a commensurabilty
 both  with and without eccentricity damping was undertaken  by \citet{LeePeale01}.
Small libration amplitudes at a similar level are attained
in both models.
The case without eccentricity damping attains higher eccentricities
and somewhat larger libration amplitudes as the evolution continues 
whereas a low steady libration amplitude is maintained in the case with eccentricity damping.
The above situation is consistent with the initial reduction in libration amplitude
that occurs prior to the attainment of steady eccentricities as not being due to
eccentricity damping but the  operation of adiabatic invariance \citep{Peale76}.}

\changed{ However, a study of simulations of the GJ876 system without stochastic
 forcing reveals that eccentricty damping does play a role in damping the
slow mode  (associated with apsidal corotation) on the eccentricity damping timescale,
whereas the fast mode appears to be only marginally  affected at least at small amplitudes.
Thus the residual librations in these cases consist  mainly if not entirely  of the fast mode.
Therefore the inclusion of eccentricity damping does impart some stability
to the commensurability through control of the slow mode and  thus  its possible interactions 
with the fast mode. 
As the slow mode involves $\zeta$ and $e_1,$
(see figure \ref{fig:twoplanets}), it is not surprising that this is affected by eccentricity
damping.}

 \changed{Note too that because the migration process itself causes libration amplitude
reduction on formation of the commensurability some stability is also implied
towards some types of disturbance that would take the system away from resonance.}

\subsection{Model 1}
The planets are initially in circular orbits at radii $a_1=4$~AU and $a_2=2$~AU.
Stochastic forcing is applied to the outer planet only
with the  diffusion coefficient  $D=6.4\cdot 10^{-3} \text{cm}^2/\text{s}^3.$
Note that  although this value is 640~times larger than the scaling given by equation (\ref{eq:scalingd}), corresponding  to a force that is 25~times larger than the simple estimate given  in section \ref{sec:scaling}, corresponding to smaller reduction factors 
resulting from a gap or dead zone, 
it
is too small for the angle $\phi_1$ to be  brought to circulation
during our runs. The results given above and summarized in figure \ref{fig:lifetime1},
and other tests,
indicate that similar results would be obtained if $D$ is reduced, but on a longer time
scale $\propto 1/D,$ provided the  migration rate is also appropriately ultimately  reduced
so that the system can survive for long enough to  enable $\zeta$ to be driven
into circulation.

The outer planet is made to  migrate inwards on a timescale $\tau_{a_1}=8000$~years. The inner planet migrated slowly outwards on a timescaly of $\tau_{a_2}=-20000$~years. This results in convergent migration.
 For both planets we use an eccentricity to semi major axis damping ratio of $K=8$. 
The resulting evolutionary timescales are significantly larger than those used by eg. \cite{SandorKley06} and more easily justified by hydrodynamical simulations.

The time evolution is shown in the left plot of figure \ref{fig:hd128311}. After resonance capturing all angles are either initially librating 
or on the border between libration and circulation.
The slow mode has a period of $\sim 700$~years, whereas the slow mode period is 20~times shorter.

It can be seen from 
figure \ref{fig:hd128311} and other related figures below, that while migration continues  libration ampliudes tend to be  controlled apart from when
$e_1$ becomes either zero or very close to zero.  
 Then, due to stochastic forcing, $\phi_2$ and $\zeta$
 start circulating. Subsequently  libration is recovered over time
intervals  for which  $e_1$
does not attain very small values but eventually additional stochastic forcing
together with the repeated attainment of small values for $e_1$
causes $\phi_2$
  to remain circulating for the remainder of the simulation.
After $13000$~years, both the forces due to migration and turbulence are reduced smoothly on a timescale of $2000$~years.  
The result is a stable configuration that resembles the observed system very well. 

\subsection{Model 2}
For this simulation, we lengthened the migration timescales by a 
factor of 2 to show that the results are generic.
 We also decreased the value of $K$ to $5.5$. 
This results in larger eccentricities and the final state
 better  resembled our representation of  the observed system. However,
it should be  kept in mind that the eccentricities
 are not well constrained by the observations. 
All other parameters are the same as for model 1. 
The time evolution is plotted in figure \ref{fig:hd128311_2}.
 The resonant angles librate immediately after the capture into
 resonance as predicted for sufficiently slow migration.
 However, in the same way as for model 1 described above,  stochastic forces
 make $\phi_2$ circulate soon afterwards. 
Once the migration forces
 and stochastic forces  are removed between $20000$ and $22000$~years, 
the system stays in a stable configuration with no apsidal corotation.

\subsection{Model 3}
All parameters are the same as for model 2. Accordingly
model 3 represents another statistical realization of that case.
The time evolution is plotted in figure \ref{fig:hd128311_3}. 
The final state is very close to the  boundary between
 libration and circulation of $\zeta.$
As discussed above, large uncertainties
in the orbital parameters  means that
the actual system could be in such a state.

Due to stochastic  forces, which may result from local turbulence, 
we are able to generate a broad spectrum of model systems.
 Some of them  undergo a strong scattering during the migration process. 
However, the surving systems resemble the observed system very well. 
Although the orbital parameters are not well constrained yet, 
we showed that values in the right range
  are naturally produced by a turbulent disk.

\section{Discussion}
\label{sec:discussion}

In this paper, we have presented a self consistent analytic model 
applicable to  either a single planet or  two planets in a mean motion resonance 
subject to external  stochastic forcing. 
The stochastic forces could result from MRI driven turbulence 
within the protoplanetary disk but our treatment is equally applicable
to any other source.

We considered the evolution of a stochasticly forced two planet system 
that is initially  deep inside a MMR  (ie. the two
independent  resonant angles librate with small amplitude).
Stochastic forces cause libration amplitudes to increase
in the mean with time until all  resonant angles  are driven into
circulation at which point the commensurability is lost.
 Often a strong scattering occurs soon afterwards for sytems
composed of planets in the Jovian mass range.

We isolated a fast libration mode 
which is associated with oscillations of the semi-major axes and a slow libration mode  which is mostly associated with the
motion of the angle between the apsidal lines of the two planets.
These modes respond differently to stochastic forcing,
 the slow mode  being more readily converted to circulation
 than the fast mode. 
This slow mode is sensitive to the attainment of small eccentricities which cause
rapid transitions \changed{between} libration and circulation.
The amplitude of the fast mode grows more regularly in the mean,
with the square of the libration amplitude in most cases increasing 
linearly with time
and being proportional to the diffusion coefficient $D$.   
Of course this discussion is simplified and there are limitations. For example
if the total mass of the system is reduced,  the disruption time eventually becomes comparable to the libration period. In that case the averaging process that we used in the derivation is no longer valid and the lifetime no longer scales as $1/D.$ 

The analytic model was compared with numerical simulations which incorporated
stochastic forces.  Those forces, parameterized
by the mean square values of each component in cylindrical ccoordinates
and the autocorrelation time,  were applied in a continuous manner giving results
 that could be directly compared with the analytic model.
The simulations were in broad agreement with
analytic  predictions and we presented illustrative examples of the disruption process. 
We performed the simulations for a large range of  diffusion parameters, planet masses and initial eccentricities to verify the scaling law for the
commensurability disruption time
summarized hereafter.

To summarize our results, recalling  that the slow angle is driven into circulation
before the fast angle,  so that the ultimate lifetime is determined
by the time taken for the fast angle to achieve circulation, we determine the lifetime, $t_f,$
using equation (\ref{eq:growthf})  setting $t=t_f,$  $(\Delta Q)^2 = (\Delta Q)^2_0 = 4,$
together with $\gamma_f =p =1,$ so obtaining
\begin{eqnarray}
 t_f&=&\frac {a_1^2 \omega_{lf}^2(\Delta Q)^2_0}{36D}.
\end{eqnarray}
We showed  above that this gives good agreement with our numerical results.
Using $D=2\langle F_i^2\rangle \tau_c,$ we can express this result in terms 
of the relative magnitude of the stochastic forcing in the form
\begin{eqnarray}
 t_f&=&2.4\times 10^{-4}\left(\frac {a_1^2n_1^4}{\langle F_i^2\rangle}\right)\nonumber
 \left(\frac{(\Delta Q)^2_0}{8n_1\tau_c}\right)\\
 && \;\;\cdot
  \left( \frac{8.5\omega_{lf}\sqrt{q_{GJ}}}{n_1\sqrt{q}}\right)^2\frac{q}{q_{GJ}} P_1,\label{RESEQ}
\end{eqnarray}
where $P_1$ is the  orbital period  of the outer planet.
Here the first quantity in brackets represents the ratio of the square of the central force 
per unit mass to the mean 
square stochastic force per unit mass acting on the outer planet.
The other quantities in brackets, scaled to the GJ876 system
 are expected to be unity, while the last factor 
$q/q_{GJ}$ is the ratio of the total mass ratio of the system
to the same quantity  for GJ876. Here it is assumed that the two planets in the system
have comparable masses.

From (\ref{RESEQ}) we see that a  non migrating system such as GJ876
could survive in resonance for  $t_f \sim 10^6$~years if the stochastic force amplitude
is $\sim 10^{-5}$ times the central force. This expression enables scaling
to other systems at other disk locations for other stochastic forcing amplitudes.
Inference of survival probabilities for particular systems depends on many 
uncertain aspects, such as the protoplanetary disk model and the strength
of MRI turbulence. 
However, the mass ratio dependence in equation (\ref{RESEQ}) indicates that survival is favored for more massive systems. 
At  the present time the number of observed
resonant systems is too small for definitive conclusions to be made.
However, the fact that several systems
exhibiting commensurabilities  have been observed
indicates that resonances are not always completely disrupted by stochastic forces due to
turbulence, 
but rather may be  modified  as in our study of HD128311.

  The HD128311 system 
is such that the fast mode librates with the slow mode being 
near the borderline between libration and circulation.
We found that such a configuration was readily produced in a scenario
in which the commensurability was formed through a temporary period of convergent migration
with the addition of stochastic forcing.  
During a migration phase  moderate 
adiabatic invariance applied to the libration modes together with eccentricity 
damping  leads to increased stabilization and a  longer lifetime for the resonance.
However, as discussed above, the time evolution of the eccentricity
and in particular the attainment of small values plays an important role
in causing the slow mode to circulate corresponding
to the loss of apsidal corotation. 
Thus we expect that a large eccentricity damping rate does not necessarily stabilize the apsidal 
corotation of the system. 
Additional simulations have  shown that this   is lost more easily for large damping rates  ($K\gg 10$).
\changed{On the other hand it should be noted that this  has less significance when one of the orbits
becomes nearly circular.}

Further observations of extrasolar planetary systems leading to better statistics may
lead to an improved situation for assessing the role
of stochastic forcing in future.

\begin{acknowledgements} 
We thank an anonymous referee for useful comments and suggestions.
HR was supported by an Isaac Newton studentship, the award of a STFC studentship and St John's College Cambridge. Numerical simulations were performed on the Hyades cluster. 
\end{acknowledgements}

\bibliography{full}

\begin{thebibliography}{30}
\expandafter\ifx\csname natexlab\endcsname\relax\def\natexlab#1{#1}\fi

\bibitem[{{Adams} {et~al.}(2008){Adams}, {Laughlin}, \& {Bloch}}]{Adams08}
{Adams}, F.~C., {Laughlin}, G., \& {Bloch}, A.~M. 2008, \apj, 683, 1117

\bibitem[{{Brouwer} \& {Clemence}(1961)}]{BrouwerClemence1961}
{Brouwer}, D. \& {Clemence}, G.~M. 1961, {Methods of celestial mechanics} (New
  York: Academic Press, 1961)

\bibitem[{{Crida} {et~al.}(2008){Crida}, {S{\'a}ndor}, \& {Kley}}]{Crida08}
{Crida}, A., {S{\'a}ndor}, Z., \& {Kley}, W. 2008, \aap, 483, 325

\bibitem[{{de Val-Borro} {et~al.}(2006){de Val-Borro}, {Edgar}, {Artymowicz},
  {Ciecielag}, {Cresswell}, {D'Angelo}, {Delgado-Donate}, {Dirksen}, {Fromang},
  {Gawryszczak}, {Klahr}, {Kley}, {Lyra}, {Masset}, {Mellema}, {Nelson},
  {Paardekooper}, {Peplinski}, {Pierens}, {Plewa}, {Rice}, {Sch{\"a}fer}, \&
  {Speith}}]{DiscComp2006}
{de Val-Borro}, M., {Edgar}, R.~G., {Artymowicz}, P., {et~al.} 2006, \mnras,
  370, 529

\bibitem[{{Fromang} \& {Papaloizou}(2007)}]{FromangPapaloizou07}
{Fromang}, S. \& {Papaloizou}, J. 2007, \aap, 476, 1113

\bibitem[{{Fromang} {et~al.}(2007){Fromang}, {Papaloizou}, {Lesur}, \&
  {Heinemann}}]{Fromangetal07}
{Fromang}, S., {Papaloizou}, J., {Lesur}, G., \& {Heinemann}, T. 2007, \aap,
  476, 1123

\bibitem[{{Gammie}(1996)}]{Gammie96}
{Gammie}, C.~F. 1996, \apj, 457, 355

\bibitem[{{Hamilton}(1994)}]{Hamilton94}
{Hamilton}, D.~P. 1994, Icarus, 109, 221

\bibitem[{{Kasdin}(1995)}]{KASDIN}
{Kasdin}, N.~J. 1995, Proceedings of the IEEE, 83

\bibitem[{{Lecoanet} {et~al.}(2008){Lecoanet}, {Adams}, \&
  {Bloch}}]{LecoanetAdams2008}
{Lecoanet}, D., {Adams}, F.~C., \& {Bloch}, A.~M. 2008, ArXiv e-prints

\bibitem[{{Lee} \& {Peale}(2001)}]{LeePeale01}
{Lee}, M.~H. \& {Peale}, S.~J. 2001, in Bulletin of the American Astronomical
  Society, Vol.~33, Bulletin of the American Astronomical Society, 1198

\bibitem[{{Lee} \& {Peale}(2002)}]{LeePeale2002}
{Lee}, M.~H. \& {Peale}, S.~J. 2002, \apj, 567, 596

\bibitem[{{Nelson}(2005)}]{Nelson2005}
{Nelson}, R.~P. 2005, \aap, 443, 1067

\bibitem[{{Nelson} \& {Papaloizou}(2004)}]{NelsonPapaloizou04}
{Nelson}, R.~P. \& {Papaloizou}, J.~C.~B. 2004, \mnras, 350, 849

\bibitem[{{Oishi} {et~al.}(2007){Oishi}, {Mac Low}, \& {Menou}}]{oishi2007}
{Oishi}, J.~S., {Mac Low}, M.-M., \& {Menou}, K. 2007, \apj, 670, 805

\bibitem[{{Papaloizou}(2003)}]{PapaloizouResonances}
{Papaloizou}, J.~C.~B. 2003, Celestial Mechanics and Dynamical Astronomy, 87,
  53

\bibitem[{{Papaloizou} \& {Szuszkiewicz}(2005)}]{PapaloizouSzuszkiewicz2005}
{Papaloizou}, J.~C.~B. \& {Szuszkiewicz}, E. 2005, \mnras, 363, 153

\bibitem[{{Papaloizou} \& {Terquem}(2006)}]{papaloizouterquem06}
{Papaloizou}, J.~C.~B. \& {Terquem}, C. 2006, Reports of Progress in Physics,
  69, 119

\bibitem[{{Peale}(1976)}]{Peale76}
{Peale}, S.~J. 1976, \araa, 14, 215

\bibitem[{Press {et~al.}(1992)}]{nr}
Press, W. {et~al.} 1992, {Numerical recipes in C} (Cambridge University Press
  Cambridge)

\bibitem[{{Reipurth} {et~al.}(2007){Reipurth}, {Jewitt}, \&
  {Keil}}]{Reipurth07}
{Reipurth}, B., {Jewitt}, D., \& {Keil}, K., eds. 2007, {Protostars and Planets
  V}

\bibitem[{{Rivera} {et~al.}(2005){Rivera}, {Lissauer}, {Butler}, {Marcy},
  {Vogt}, {Fischer}, {Brown}, {Laughlin}, \& {Henry}}]{Rivera2005}
{Rivera}, E.~J., {Lissauer}, J.~J., {Butler}, R.~P., {et~al.} 2005, \apj, 634,
  625

\bibitem[{{S{\'a}ndor} \& {Kley}(2006)}]{SandorKley06}
{S{\'a}ndor}, Z. \& {Kley}, W. 2006, \aap, 451, L31

\bibitem[{Schneider(2009)}]{exoplanet}
Schneider, J. 2009, \texttt{http://exoplanet.eu}

\bibitem[{{Sinclair}(1975)}]{Sinclair1975}
{Sinclair}, A.~T. 1975, \mnras, 171, 59

\bibitem[{{Snellgrove} {et~al.}(2001){Snellgrove}, {Papaloizou}, \&
  {Nelson}}]{SnellgrovePapaloizouNelson01}
{Snellgrove}, M.~D., {Papaloizou}, J.~C.~B., \& {Nelson}, R.~P. 2001, \aap,
  374, 1092

\bibitem[{Stoer \& Bulirsch(2002)}]{StoerBulirsch02}
Stoer, J. \& Bulirsch, R. 2002, {Introduction to Numerical Analysis} (Springer)

\bibitem[{{Udry} {et~al.}(2007){Udry}, {Fischer}, \& {Queloz}}]{Udry07}
{Udry}, S., {Fischer}, D., \& {Queloz}, D. 2007, in Protostars and Planets V,
  ed. B.~{Reipurth}, D.~{Jewitt}, \& K.~{Keil}, 685--699

\bibitem[{{Vogt} {et~al.}(2005){Vogt}, {Butler}, {Marcy}, {Fischer}, {Henry},
  {Laughlin}, {Wright}, \& {Johnson}}]{Vogt2005}
{Vogt}, S.~S., {Butler}, R.~P., {Marcy}, G.~W., {et~al.} 2005, \apj, 632, 638

\bibitem[{{Weidenschilling}(1977)}]{Weidenschilling1977}
{Weidenschilling}, S.~J. 1977, \apss, 51, 153

\end{thebibliography}
\bibliographystyle{aa}

\end{document}